\documentclass[twocolumn,prb,showpacs,preprintnumbers,superscriptaddress,amsmath,amssymb,citeautoscript,10pt,aps]{revtex4-1}
\usepackage{graphicx}
\usepackage{dcolumn}
\usepackage{color}
\usepackage{bm}
\usepackage[hidelinks]{hyperref}
\hypersetup{colorlinks,citecolor=blue,filecolor=blue,linkcolor=blue,urlcolor=blue}

\newcommand{\ud}{\,\mathrm{d}}
\newcommand{\im}{\mathrm{i}}
\newcommand{\e}{\textrm{e}}
\newcommand{\dpsi}{\psi^\dagger}
\newcommand{\tphi}{{\tilde{\phi}}}

\DeclareMathOperator{\sgn}{sgn}

\DeclareMathOperator{\diag}{diag}

\begin{document}

\title{\texorpdfstring{Conductance in inhomogeneous quantum wires:\\ Luttinger liquid predictions and quantum Monte Carlo results}{Conductance in inhomogeneous quantum wires: Luttinger liquid predictions and quantum Monte Carlo results}}

\author{D. Morath}
\email{dmorath@physik.uni-kl.de}
\affiliation{Department of Physics and Research Center OPTIMAS, University of Kaiserslautern, D-67663 Kaiserslautern, Germany}
\author{N.~Sedlmayr}
\email{nsedlmayr@hotmail.com}
\affiliation{Department of Physics and Astronomy, Michigan State University, East Lansing, Michigan 48824, USA}
\author{J.~Sirker}
\affiliation{Department of Physics and Astronomy, University of Manitoba, Winnipeg R3T 2N2, Canada}
\author{S.~Eggert}
\affiliation{Department of Physics and Research Center OPTIMAS, University of Kaiserslautern, D-67663 Kaiserslautern, Germany}

\date{\today}

\begin{abstract}
We study electron and spin transport in interacting quantum wires
contacted by noninteracting leads. We theoretically model the wire and
junctions as an inhomogeneous chain where the parameters at the
junction change on the scale of the lattice spacing. We study such
systems analytically in the appropriate limits based on Luttinger
liquid theory and compare the results to quantum Monte Carlo
calculations of the conductances and local densities near the
junction. We first consider an inhomogeneous spinless fermion model
with a nearest-neighbor interaction and then generalize our results to
a spinful model with an onsite Hubbard interaction.
\end{abstract}


\maketitle

\section{Introduction}

\begin{figure}[!ht]
\includegraphics[width=0.99\columnwidth]{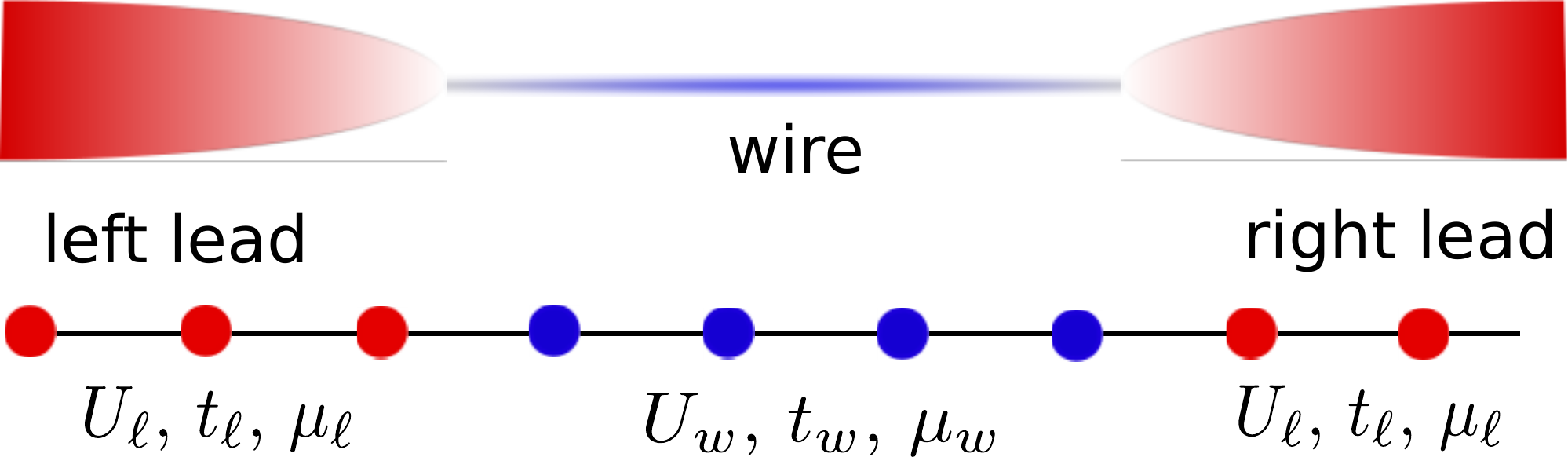} 
\caption{(Color online) A quantum wire connected to two noninteracting leads. 
The identical leads are modeled as a chain with hopping $t_\ell$, chemical potential $\mu_\ell$, and interaction $U_\ell=0$. 
The wire has parameters $U_w,\, t_w,\, \mu_w$. The junction between the leads and the wire is modeled as being abrupt.}
\label{Fig1}
\end{figure}

An important tool to study the physics of quantum wires is
measurements of their conductance as a function of parameters such as
the filling fraction or
temperature.\cite{Liang2001,Javey2003,Yacoby1996,Steinberg2008,Tarucha1995,Purewal2007,KamataKumada}
In order to understand the results of such experiments it is important
to find an appropriate model not only for the quantum wire itself but
for the full system including the leads. Typically, the properties of
the quantum wire are strongly affected by electron-electron
interactions. Fermi liquid theory has to be replaced by Luttinger
liquid theory in one
dimension.\cite{Tomonaga1950,Luttinger1963,Giamarchi2004} The leads,
on the other hand, form a higher-dimensional electron gas in which
interactions can be neglected. This suggests that a lead-wire-lead
system can be modeled as an inhomogeneous quantum wire where the
interaction and hopping parameters, as well as the chemical potential,
change at the junctions. A sketch of such a setup and how it is
modeled as an inhomogeneous wire is shown in Fig.~\ref{Fig1}. Quantum
wires have been analyzed using Luttinger liquid theory previously and
it has been shown that for perfect adiabatic contacts the conductance
of the wire is controlled by the parameters of the lead rather than of
the
wire\cite{Yue1994,Safi1995,Maslov1995,Ogata1994,Wong1994,Chamon1997,Safi1999,Imura2002,Enss2005,*Janzen2006,Rech2008,Rech2008a,Gutman2010a,Thomale2011,Sedlmayr2012a,Sedlmayr2013,Sedlmayr2014a}. The
conductance for adiabatic contacts with noninteracting leads is
therefore given by the perfect quantum conductance, $G=ne^2/h$ for $n$
channels, instead of being renormalized by the Luttinger liquid of the
wire as might be expected from a naive calculation for an infinite
wire. However, for any reasonably sharp junction there will be
scattering at the junction even for otherwise perfect ballistic
connections. Such scattering becomes renormalized by the interaction
and can lead to a vanishing d.c.~conductance in the low temperature
limit for repulsive
interactions.\cite{Kane1992,Kane1992a,Eggert1992,Furusaki1993,Furusaki1996,Pereira2004a,Sedlmayr2011b}

In two recent papers\cite{Sedlmayr2012a,Sedlmayr2014a} 
we have shown
that for an inhomogeneous spinless fermion model as depicted in
Fig.~\ref{Fig1} it is, however, still possible to obtain perfect
conductance by tuning the parameters of the wire and the leads.  
Using Luttinger liquid calculations and a comparison with
numerical quantum Monte Carlo (QMC) results for static local response functions 
it was possible to establish the existence of 
a highly nontrivial conducting fixed point described 
by two effective Luttinger liquid parameters.\cite{Sedlmayr2012a,Sedlmayr2014a}  
Here, we want to generalize these studies
in two ways. First, we will check the existence of conducting fixed
points more directly by calculating the conductance numerically using
QMC. Second, we will generalize the study of the conductance in
inhomogeneous wires to spinful systems. We will concentrate on two
microscopic models: (1) The spinless fermionic chain with Hamiltonian
$H=H_0+H_I$ where
\begin{eqnarray}
\label{XXZ}
H_0 &=& -\sum_j \left(t_j \Psi_{j+1}^\dagger\Psi_j+\textrm{H.c.}+\mu_j n_j\right)\,,\textrm{ and}\nonumber \\
H_I &=& \sum_j U_j n_j n_{j+1}\,.
\end{eqnarray}
Here $\Psi_j^{(\dagger)}$ is the annihilation (creation) operator of a
spinless fermion at site $j$ and $n_j=\Psi_{j}^\dagger\Psi_j$ is the
density operator. The site-dependent parameters $t_j,\mu_j$ and $U_j$
are defined as shown in Fig.~\ref{Fig1}. (2) The inhomogeneous Hubbard
model:
\begin{eqnarray}
\label{Hubbard}
H_0 &=& -\sum_{j,\sigma} \big(t_j \Psi_{j+1,\sigma}^\dagger\Psi_{j,\sigma}+\textrm{H.c.}+\mu_j n_j\big)\,,\textrm{ and}\nonumber \\
%% &&\hspace{1cm}+\mu_BB_j(n_{j,\uparrow}-n_{j,\downarrow})\big)\,,\textrm{ and}\nonumber \\
H_I &=& \sum_j U_j n_{j,\uparrow} n_{j,\downarrow}\,,
\end{eqnarray}
where $\Psi_{j,\sigma}^{(\dagger)}$ is now the annihilation (creation)
operator of an electron with spin $\sigma$. The particle number for
each spin species is given by
$n_{j,\sigma}=\Psi_{j,\sigma}^\dagger\Psi_{j,\sigma}$ and the total
number operator is $n_{j}=n_{j,\uparrow}+n_{j,\downarrow}$. For the
numerical simulations we will consider systems with periodic boundary
conditions with half of the system representing the noninteracting
leads and the other half the interacting quantum wire. It is important
to note that the backscattering at the two junctions will not
influence each other as long as we ensure that the distance between
the junctions is large compared to the correlation length in the
quantum wire, $\xi\sim u/T$, where $u$ is the velocity of elementary
excitations and $T$ the temperature.

Our paper is organized as follows. In Sec.~\ref{QMCSec} we will
introduce the QMC method used to calculate the conductance and discuss
cases of homogeneous and inhomogeneous wires where exact results are
available which can be used to check the accuracy of the numerical
results. In Sec.~\ref{XXZSec} we then present results for the spinless
fermionic chain, Eq.~\eqref{XXZ}. Next, we derive the bosonized theory
for the inhomogeneous Hubbard chain in Sec.~\ref{analyticSec} and
compare the theoretical predictions with QMC data. We summarize our
main results and discuss some of the remaining open questions in
Sec.~\ref{sec_conclusion}.

\section{QMC method}\label{QMCSec}

We have implemented a quantum Monte Carlo (QMC) algorithm, the stochastic
series expansion (SSE),\cite{Sandvik1992} to calculate imaginary time
correlation functions.\cite{Dorneich2001} The conductance of the wire
in linear response can then be obtained from these imaginary time
correlation functions.\cite{Louis2003}
We calculate the linear response to an infinitesimal drop
in electric and magnetic field at site $k$ for charge and spin
respectively
\begin{eqnarray}
\label{drops}
P_{k}^{c}&=& e\sum\limits_{m>k}(n_{m,\uparrow}+n_{m,\downarrow})\\
\nonumber P_{k}^{s}&=& \frac{\mu_{\text{B}}}{2}\sum\limits_{m>k}(n_{m,\uparrow}-n_{m,\downarrow})\,,
\end{eqnarray}
where $e$ is the elementary charge, $m$ is the site, and $\mu_{\text{B}}$ the Bohr magneton. Accordingly,
 we define a local charge and spin current operator
\begin{eqnarray}
\label{currents}
j_k^c &= &\im t_k e\sum\limits_{\sigma}({\psi}^{\dagger}_{k,\sigma}{\psi}_{k+1,\sigma}-{\psi}^{\dagger}_{k+1,\sigma}{\psi}_{k,\sigma})\nonumber \\
j_k^s &= &\frac{\im t_k\mu_{\text{B}}}{2}\left[({\psi}^{\dagger}_{k,\uparrow}{\psi}_{k+1,\uparrow}-{\psi}^{\dagger}_{k+1,\uparrow}{\psi}_{k,\uparrow})\right.\\
&&\hspace{1cm}-\left.({\psi}^{\dagger}_{k,\downarrow}{\psi}_{k+1,\downarrow}-{\psi}^{\dagger}_{k+1,\downarrow}{\psi}_{k,\downarrow})\right]\,.\nonumber
\end{eqnarray}
Following Ref.~\onlinecite{Louis2003} we calculate the charge and spin conductance ($\nu=c,s$) in linear response using
\begin{eqnarray}
\label{conduct1}
g^{\nu}_{x,y}(\omega_n)&=&-\frac{\omega_n}{\hbar}\,\mbox{Re}\,\int\limits_{0}^{\hbar\beta}\e^{\im\omega_n\tau}\langle P^{\nu}_xP^{\nu}_y(\im \tau)\rangle \ud\tau\,, \nonumber \\
&=& -\frac{\omega_n}{\hbar}\int\limits_{0}^{\hbar\beta}\cos(\omega_n\tau)\langle P^{\nu}_xP^{\nu}_y(\im \tau)\rangle \ud\tau
\end{eqnarray}
where $y=0$ is the location of the perturbation $P_y$ (quadratic in the
Fermi operators) and $x$ is the location where we determine the
response to that perturbation, where $|x-y|$ must be small. Here $\omega_n= 2 \pi n / \beta $ are
the bosonic Matsubara frequencies, which are used to extrapolate to
$\omega=0$ to obtain the d.c.~conductance. For the spinless fermionic
chain \eqref{XXZ} we can only define a charge conductance using a
voltage drop $P_k^c=e\sum_{m>k} n_m$. The charge current operator
$j_k^c$ is then defined as in Eq.~\eqref{currents} but without the
spin index $\sigma$.

Numerically, the task of obtaining conductances is now reduced to
calculating expectation values in imaginary time. The technique for
this is described in Ref.~\onlinecite{Dorneich2001}.  QMC provides us
with the expectation values $\langle
P^{\nu}_xP^{\nu}_y(\im\tau)\rangle$ which are periodic in $\tau$ with
a period of $\beta$. As a final step, we have to numerically perform
the integral in Eq.~\eqref{conduct1} to obtain the conductances.

\begin{figure}
\includegraphics[width=0.95\columnwidth]{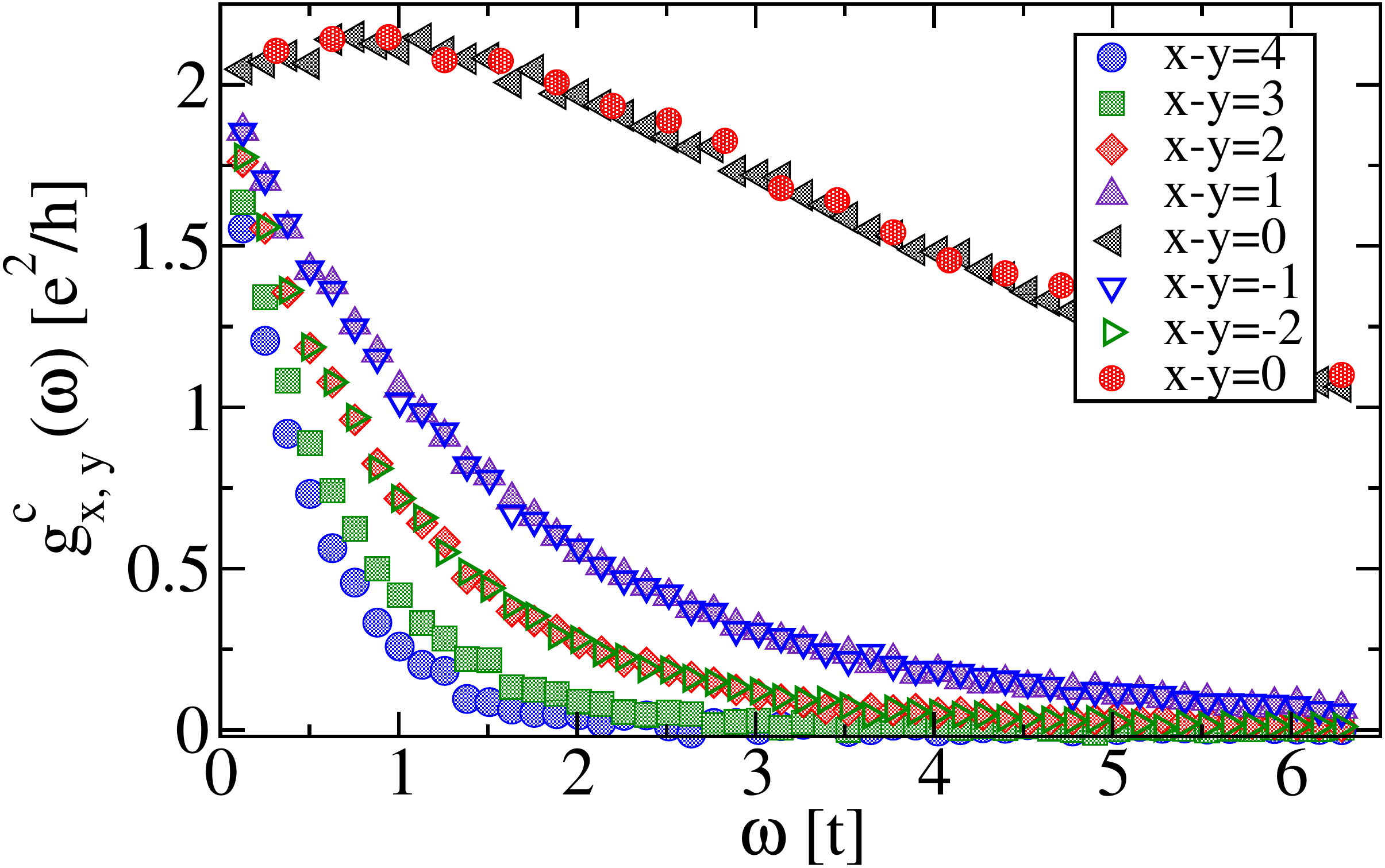} 
\caption{(Color online) The charge conductance $g^{c}_{x,y}(\omega_n)$ for a noninteracting chain of spinful fermions with length $L=200$, periodic boundary conditions, and inverse temperature $\beta t=50$. Results for different distances $x-y$ between the perturbation and the measurement are displayed, for $x-y=0$ results are also shown for $\beta t=20$ (red circles). They all extrapolate in the $\omega\to0$ limit to a conductance of $g^{c}_{x,y}(\omega\to 0)\to2e^2/h$ as expected, see main text. The errors in the data are smaller than the point size.}
\label{Fig2}
\end{figure}

\begin{figure}
\includegraphics[width=0.95\columnwidth]{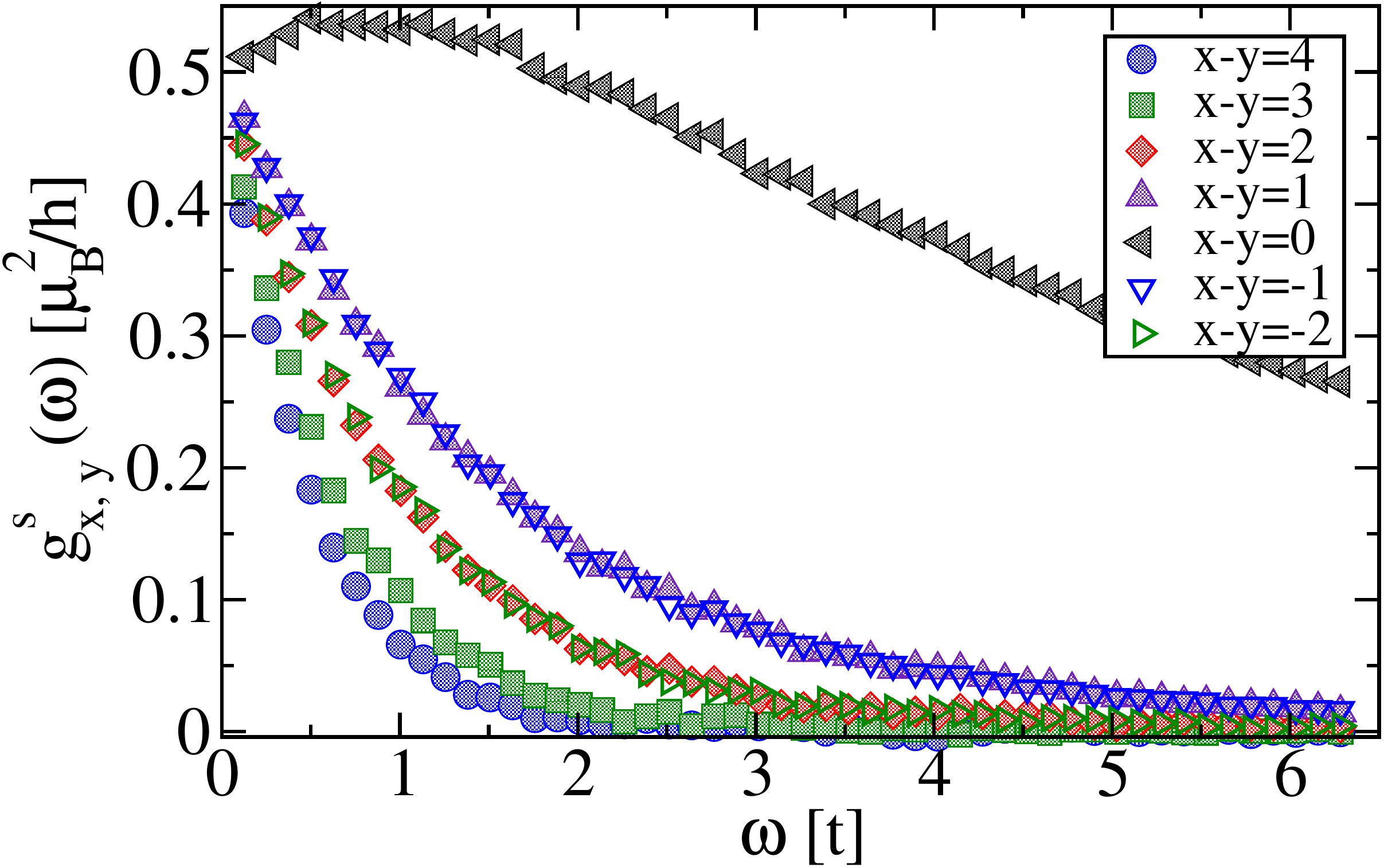} 
\caption{(Color online) The spin conductance $g^{s}_{x,y}(\omega_n)$ for the same system and parameters as in Fig.~\ref{Fig2}. All curves extrapolate in the $\omega\to0$ limit to the theoretically expected ideal conductance of $g^{s}_{x,y}(\omega\to0)\to0.5\mu_B^2/h$. As for Fig.~\ref{Fig2} the errors are smaller than the point size.}
\label{Fig3}
\end{figure}

In the following we discuss several consistency checks. Here it is
important to note that while this method has been described and
applied to homogeneous chains in Ref.~\onlinecite{Louis2003} it has
never before been applied to inhomogeneous chains, which is the case we
are interested in here. As a first check of our QMC algorithm we show
results for the spin and charge conductance of a homogeneous chain of
spinful noninteracting fermions (Eq.~\eqref{Hubbard} with $U_j\equiv
0$) in Fig.~\ref{Fig2} and
Fig.~\ref{Fig3}.
Independent of the distance $x-y$ between the perturbation and the
response all curves have the same direct current (d.c.),
i.e.~$\omega\rightarrow 0$, limit. Furthermore we can see that the
conductances at finite frequencies only depend on the absolute value
$\vert x-y\vert$.  Thus we extrapolate the curves for different
distances $\vert x-y\vert$ and average their $g^\nu(\omega=0)$ values
in order to obtain the d.c.~conductance, $g^\nu$. For the
extrapolation we use a polynomial fit of degree six,
$g^{\nu}_{x,y}(\omega_n)=
g^\nu(\omega=0)+\sum_{i=1}^6C_i\omega_n^i$. The fitting procedure and
the differences in $g^\nu(\omega=0)$ for the different distances
$|x-y|$ give an estimate for the error of the numerically obtained
d.c.~conductance. It is important to stress that the errors are
completely dominated by the extrapolations. The statistical errors of
the simulations at frequencies $\omega_n$ are very small and have
almost no influence on the extrapolated value for the
d.c.~conductance. Note also that we can only provide a sensible {\it
error estimate}. The true error is unknown and might in some cases be
larger than the estimated error.

In order to ensure the junctions behave independently of each other we require $T\gg u/L$ to be satisfied. In this case the simulation results remain independent of length, so that no additional
finite size scaling is required. Therefore, the systematic extrapolation
to a vanishing Matsubara frequency will give results in the thermodynamic limit.

When we run our simulations at
higher temperatures the Matsubara frequencies are further apart from
each other, see Fig.~\ref{Fig2}, 
which makes the
extrapolation to the zero frequency limit more difficult. 
On the other hand, since SSE is a high
temperature expansion, lower temperatures will increase the simulation
time, especially because in our case  measurements of
imaginary time correlation functions are necessary for
all $\tau$, and we will require larger system lengths to satisfy $T\gg u/L$. It turns out that $\beta t =50$ is a good compromise between
reasonable simulation times and a good accuracy of the extrapolation
$\omega\to 0$. As expected for a non-interacting system the
d.c.~conductance is perfect, i.e.~we find $2 e^2/h$ for the charge
conductance, since we have two independent charge channels
($\sigma=\uparrow,\downarrow$), see
Fig.~\ref{Fig2}. Similarly we find $0.5
\mu_{\text{B}}^2/h$ for the spin conductance consistent with the spin being $1/2$ in
units of $\mu_{\text{B}}$, see Fig.~\ref{Fig3}.

\begin{figure}
\includegraphics[width=0.95\columnwidth]{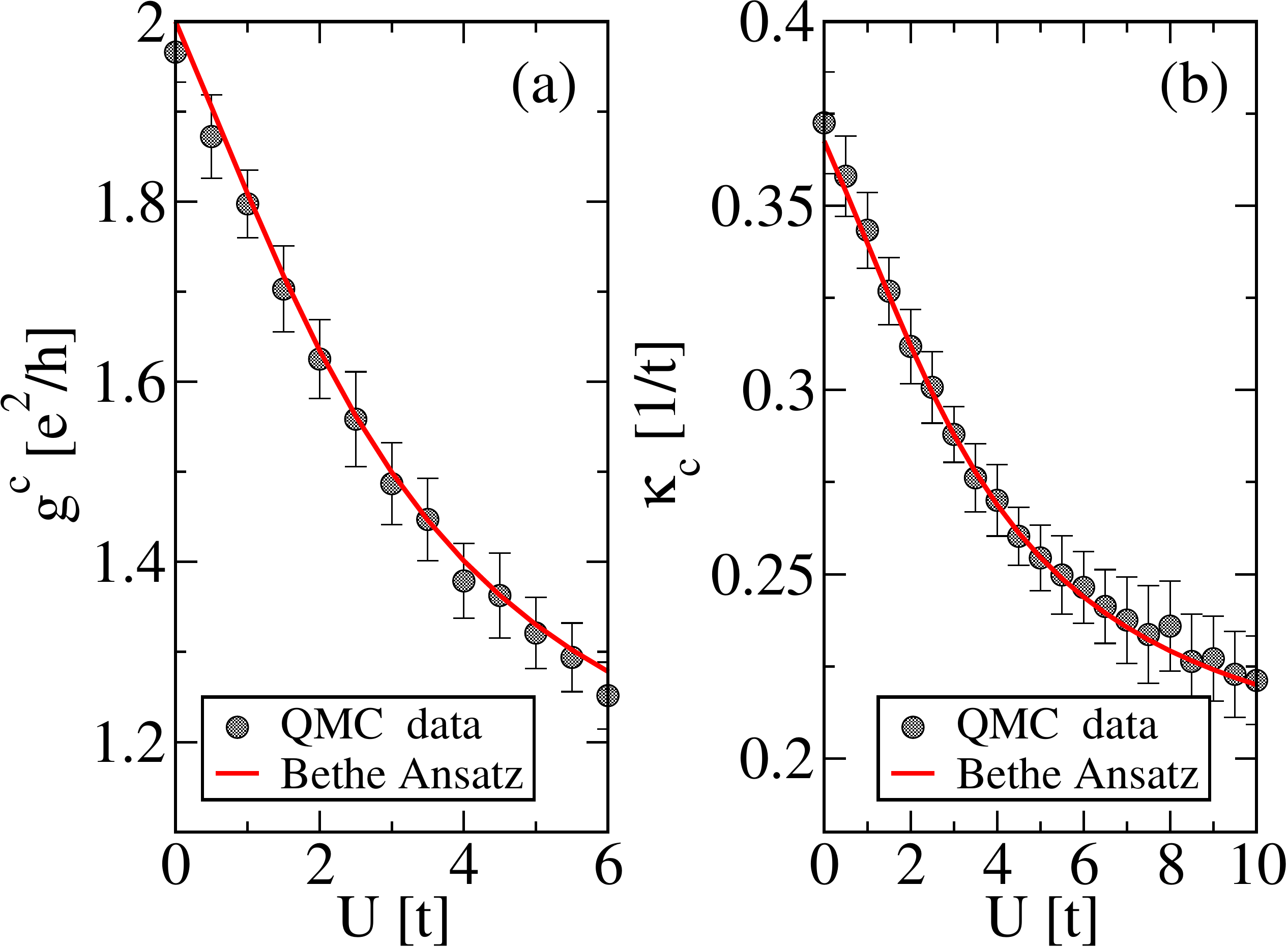} 
\caption{(Color online) Comparison between QMC data for the homogeneous Hubbard chain \eqref{Hubbard} with parameters  $\mu=t$, $\beta t =50$, and $L=150$, and the exact Bethe ansatz results. Panel (a) shows the charge conductance, panel (b) the charge compressibility.}
\label{Fig4}
\end{figure}

Next, we consider the homogeneous interacting Hubbard model which is
integrable by Bethe ansatz. In particular, the Luttinger liquid (LL)
parameters $K_\nu$ as well as the spin and charge velocities $u_{\nu}$
of the elementary excitations can be determined exactly.\cite{Essler2005} For the
conductances and compressibilities one finds, in
particular,\cite{Shirakawa2009,Sano2004,Giamarchi2004,Essler2005}
\begin{eqnarray}
\label{BA_results}
\kappa_{\nu}&=&\frac{2 K_{\nu}}{\pi u_{\nu}}\,,\,\, g^{c}=\frac{2 e^2}{h}K_{c}\,,\textrm{ and}\,\, g^{s}=\frac{\mu_{\text{B}}^2}{2h}K_{s}\,. 
\end{eqnarray}
We are considering here only the $SU(2)$ symmetric case where the spin
LL parameter is fixed, $K_s=1$. In Fig.~\ref{Fig4}(a) we
show a comparison between the QMC result for the charge conductance at
fixed chemical potential $\mu=t$ and various interaction strengths
$U$ after extrapolating to the zero frequency limit and the Bethe
ansatz result \eqref{BA_results}. To obtain the LL parameter $K_c$, an
integral equation obtained by Bethe ansatz\cite{Essler2005} has been evaluated
numerically. 
In Fig.~\ref{Fig4}(b) we show a similar comparison for the
charge compressibility. 
The QMC data in Fig.~\ref{Fig4} generally agree quite well with the exact
results for all interaction strengths $U$.

\begin{figure}
\includegraphics[width=0.95\columnwidth]{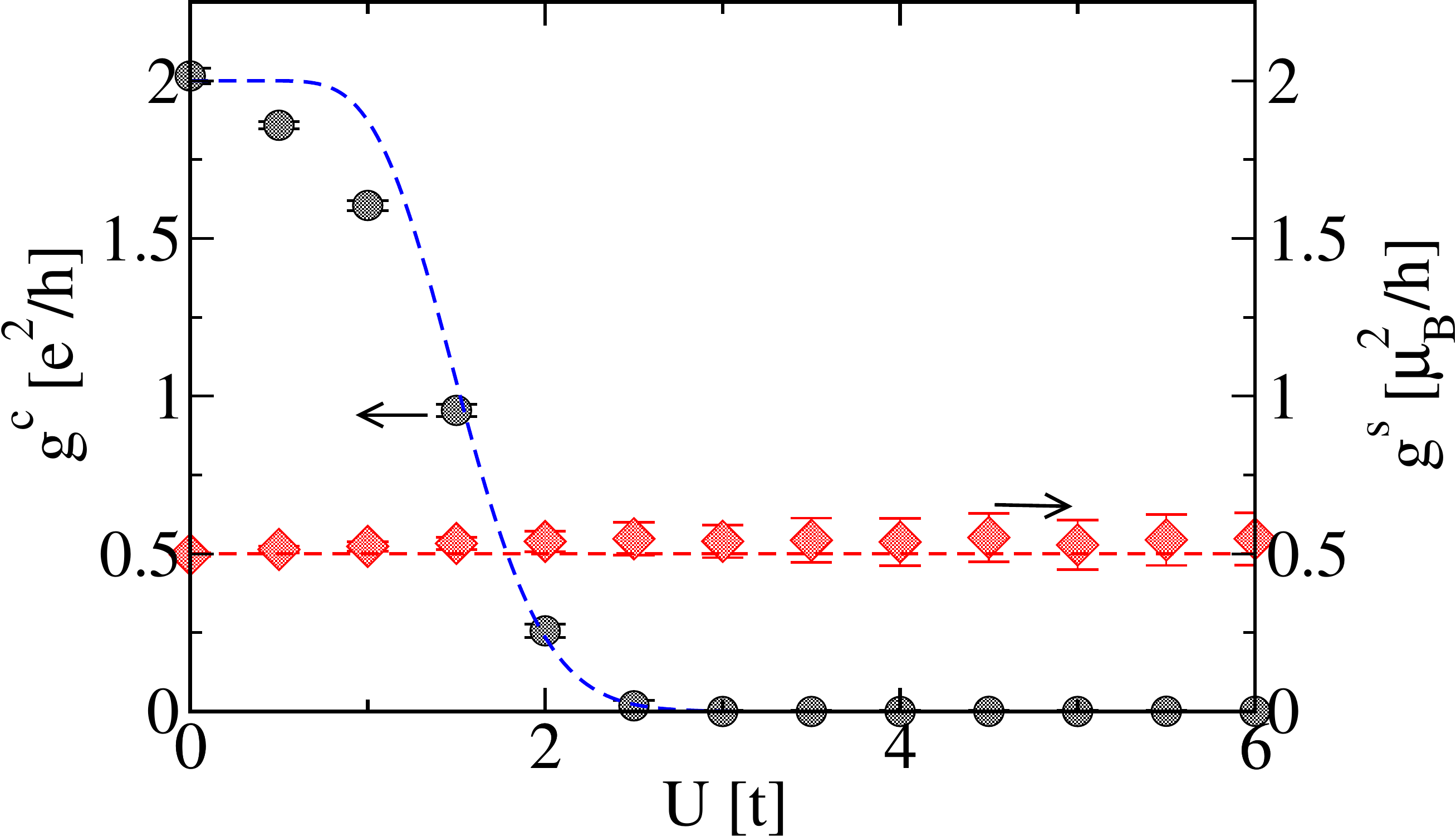} 
\caption{(Color online) The Hubbard chain at half-filling ($\mu=0$, $\beta t =50$, $L=150$). The charge conductance only drops slowly to zero because of the exponentially small charge gap at small $U$. The drop in $g^c$ is well described by Eq.~\eqref{charge_gap2} (blue dashed line). The spin conductance, on the other hand, is independent of $U$ and fixed by the $SU(2)$ symmetry (red dashed line).}
\label{Fig5}
\end{figure}

For the half-filled case, $\mu_j=0$, it is known that the Hubbard
model shows a Mott transition at arbitrarily small $U>0$ from a
conducting to an insulating ground state. The charge gap $\Delta_c(U)$, measured in units of $t$,
can be calculated by Bethe ansatz and is given by\cite{Essler2005}
\begin{eqnarray}
\label{charge_gap}
\Delta_c &=& -2+\frac{U}{2}+2\int_0^\infty \frac{d s}{s}\frac{J_1(s)\exp(-sU/4)}{\cosh(sU/4)} \nonumber \\
&\approx & \frac{4}{\pi}\sqrt{U}\exp\left(-\frac{2\pi}{U}\right)
\end{eqnarray}
where $J_1(s)$ is a Bessel function. The second line in
Eq.~\eqref{charge_gap} represents the result for small $U$ where the
charge gap is exponentially small. For large $U$, the charge gap will
scale linearly in the Hubbard interaction $U$. The spin channel, on
the other hand, remains gapless, the spin conductance is independent
of $U$, and the Luttinger parameter is fixed in the thermodynamic limit to $K_s=1$ due to the
$SU(2)$ symmetry. In the QMC data shown in Fig.~\ref{Fig5}
the spin conductance is indeed close to $g^s=\mu_B^2/2h$. Note that
for finite lengths $L$ there will be logarithmic corrections, $K_s\sim
1+1/\ln(L/L_0)$ with a characteristic length scale $L_0$,\cite{Soeffing2013} which might
partly explain why the QMC data for the spin conductance are 
slightly larger than the thermodynamic limit result. For the charge
conductance we find finite values for $U<2$ and values close to zero
for $U>2$.

To understand these results it is important to stress that
the QMC results are for finite chains of length $L=150$ at a finite temperature $T\gg u/L$. The charge
gap $\Delta_c$ leads to a characteristic temperature scale $T_c\sim
\Delta_c$ and we expect the conductance to scale as
\begin{equation}
\label{charge_gap2}
g^c(T,U)=g^c_0\exp(-T_c/T)\,,
\end{equation}
i.e.~the conductance will only become zero for temperatures small compared to $\Delta_c$. We also require chain lengths which
are large compared to $L_c(U)$, a characteristic length scale $L_c\sim
1/\Delta_c$, which will be satisfied due to the condition on the temperature and $T\gg u/L$.
Since the charge gap is exponentially
small for small $U$, very small temperatures are required to see the charge
gap in the conductance. The numerical results are well described by
setting $T_c\sim \Delta_c$ and using the small $U$ expansion for the
charge gap given by Eq.~\eqref{charge_gap}.

\begin{figure}
\includegraphics[width=0.95\columnwidth]{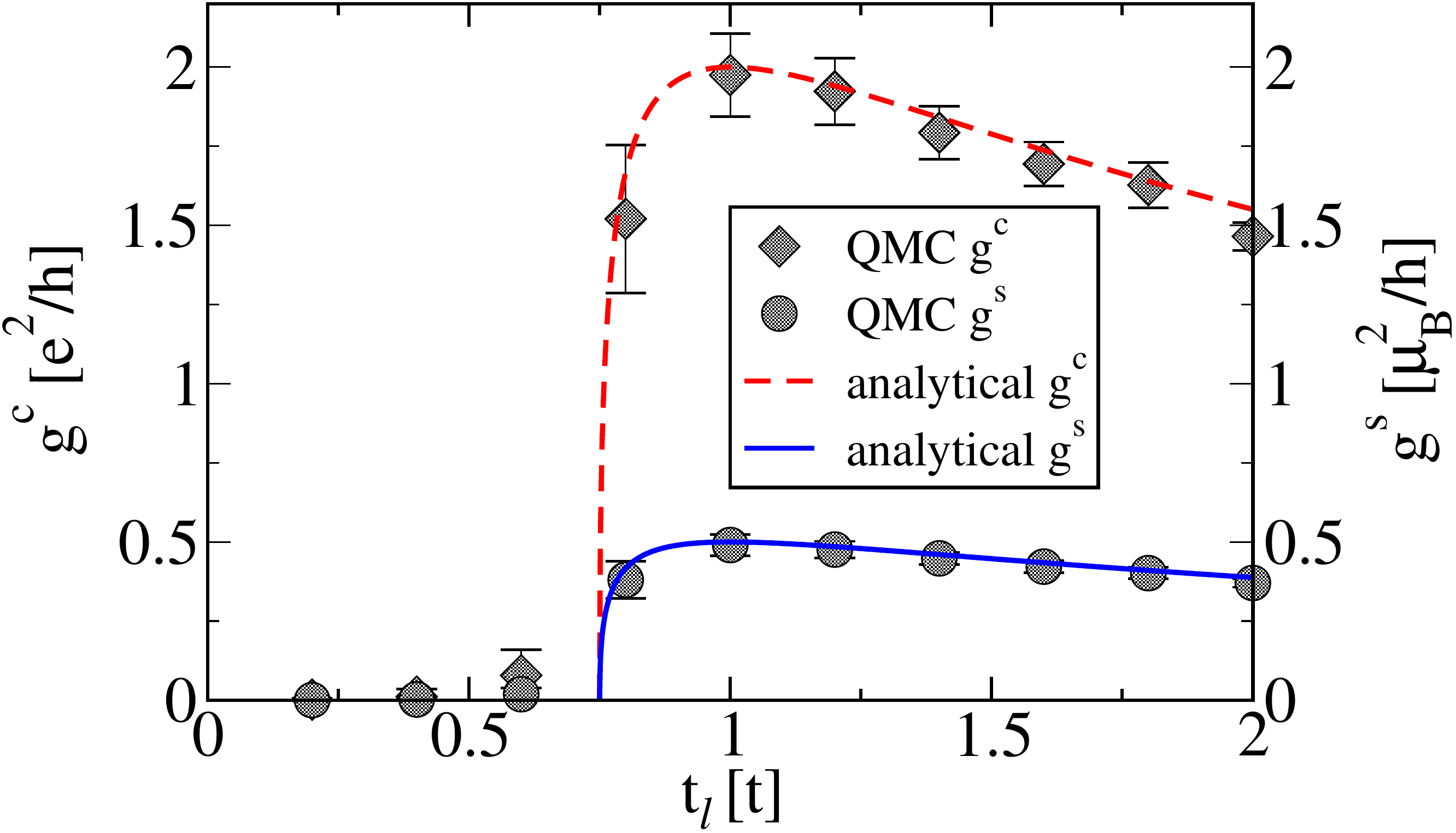} 
\caption{(Color online) QMC data for the charge and spin conductances of a junction of two non-interacting wires (symbols) with $t_w = t$, $\mu=1.5t$, $\beta t =50$, and $L=200$ compared to the exact result (lines).}
\label{Fig6}
\end{figure}

So far we have concentrated on testing the QMC algorithm for
homogeneous systems. As a next step, we consider a simple example for
a noninteracting spinful inhomogeneous system where the QMC results
can be directly compared to an analytical solution. As in all the
inhomogeneous models discussed in the following we are considering a
periodic chain of length $L$ with parameters as given in Table
\ref{table_parameters}. Here we set $U_\ell=U_w=0$ while the hopping strengths are different, $t_{\ell} \neq t_w$. 
The transmission and reflection amplitudes for non-interacting
spinless fermions are known in this case.\cite{Sedlmayr2014a} Since
the non-interacting Hubbard model has two independent spin channels,
the reflection and transmission follows directly from the spinless
result. The two velocities in the left and right part of the chain for
each spin channel are given by
\begin{equation}
u_{\ell,w;\sigma}=2at_{\ell,w}\sin\left[ak_{\ell,w;\sigma}\right]\,,
\end{equation} 
where $k_{\ell,w;\sigma}=\frac{1}{a}\arccos{\frac{\mu_\sigma}{2 t_{\ell,w}}}$
are the Fermi momenta in the lead and the wire, and $a$ is the lattice
spacing. From this the reflection coefficient can be written as
\begin{equation}
R=-\frac{u_\ell-u_w}{u_{\ell}+u_w}
\end{equation}
leading to a transmission
\begin{equation}
\vert T \vert ^ 2 = \left(1-\vert R\vert ^2\right)\frac{u_{\ell}}{u_w}\,.
\end{equation}
The conductance for each spin species is therefore given by
$G=\left(1-\vert R\vert ^2\right)e^2/h$ so that $g^c=2G$. An
analogous calculation leads to $g^s=\left(1-\vert R\vert
^2\right)\mu_B^2/2h$. These analytical results are shown as lines in
Fig.~\ref{Fig6} and compared to the QMC data.
\begin{table}
\begin{ruledtabular}
{ \renewcommand{\arraystretch}{1.5} \renewcommand{\tabcolsep}{0.7cm}
\begin{tabular}{c || c | c}  &  $ -L/2\leq j<0 $  \hspace{0.2cm}&  $0\leq j<L/2 $  \\ \hline
$t_{j}$ & $t_{\ell}=1$ & $t_{w}$ \\ \hline
$U_{j}$ & $U_{\ell}=0$ & $U_{w}>0$ \\ \hline
$\mu_{j}$ & $\mu_{\ell} $ & $\mu_{w}$ %\hline
% $B_{j}$ & $B_{\ell}$ & $B_{w}$\\ \hline 
\end{tabular}
  \caption{(Color online) \label{table_parameters} The parameters of the inhomogeneous system.}
}\end{ruledtabular}
\end{table}
As soon as both bands start to become filled the conductance increases
drastically up to a maximum at the homogeneous point and then slowly
drops down. The QMC results are in good agreement with the theoretical
prediction.

\section{Spinless inhomogeneous fermion chains}\label{XXZSec}

Here we study the interacting spinless fermion model \eqref{XXZ}. Analytically, we
have investigated this model already in two recent
publications.\cite{Sedlmayr2012a,Sedlmayr2014a} Our main result was
that there exists a line of non-trivial conducting fixed points where
the backscattering at the junction vanishes despite the inhomogeneity
of the system. In Ref.~\onlinecite{Sedlmayr2014a} we have, in
particular, been able to formulate a conformally invariant boundary
theory which describes these fixed points. One prediction of this
theory was that two different boundary Luttinger parameters exist
which determine the scaling of autocorrelations in imaginary time at
the boundary. We have been able to verify these scaling predictions
numerically by quantum Monte Carlo simulations. Furthermore, we also
obtained an analytic formula for the Friedel oscillations\cite{Friedel1958} in the
density near the boundary which are known to 
have a characteristic amplitude\cite{Egger1995,Eggert1995,Soeffing2009} 
and give information about the
interacting correlation functions and the strength of
the backscattering.\cite{Sedlmayr2012a,Sedlmayr2014a,Rommer2000a}
However, at that time we have not been able to check the main
prediction---the existence of a line of conducting fixed
points---directly. The aim of this section is to provide such a direct
check using the QMC method described in the previous section.

\subsection{The half-filled case}
\begin{figure}
\includegraphics[width=0.95\columnwidth]{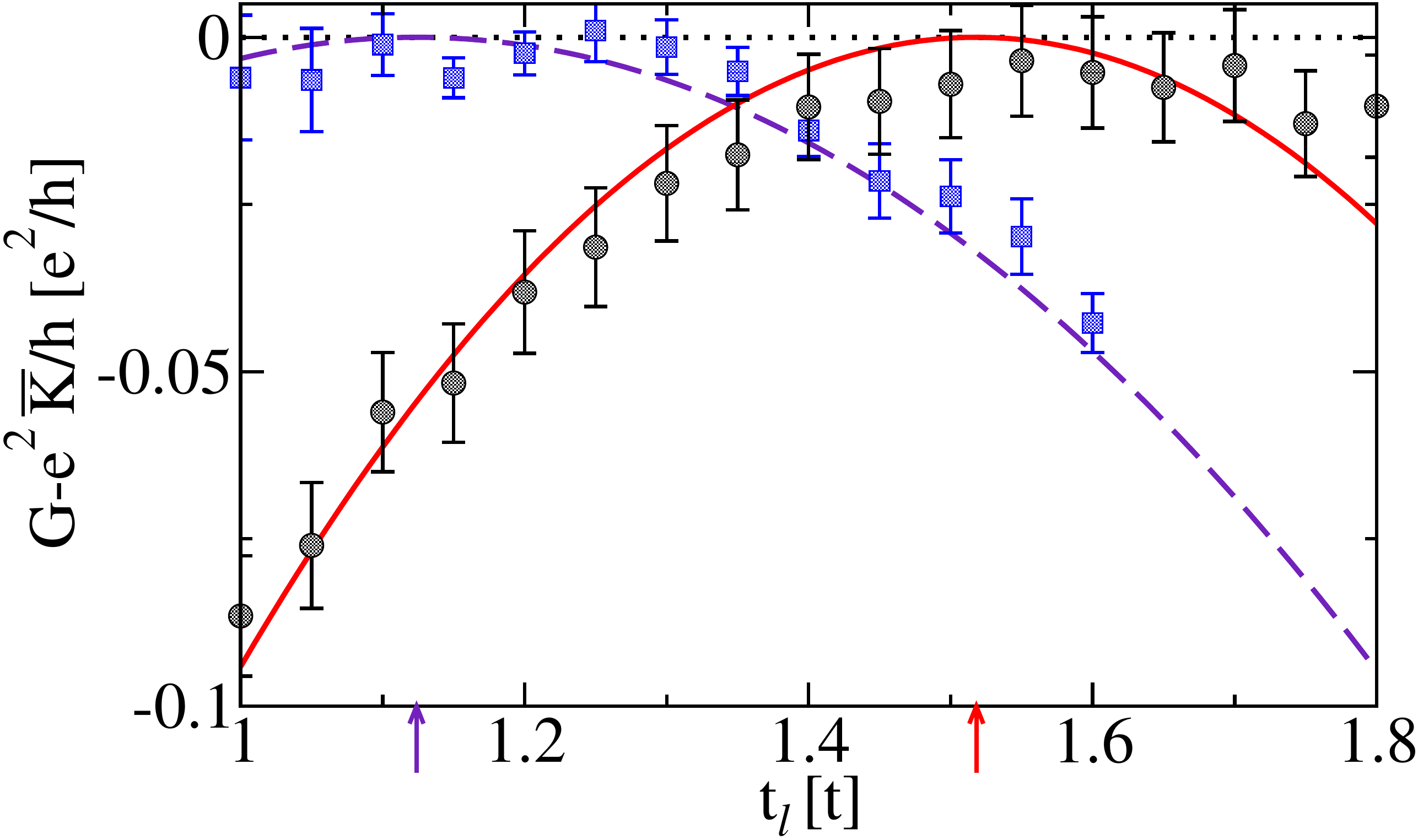} 
\caption{(Color online) Comparison between QMC data (symbols) for the inhomogeneous spinless chain \eqref{XXZ} with parameters  $\mu=0$, $t_w=t$, $U_\ell=0$, $\beta t =50$, $U_w=1.8t$ (black), $U_w=0.4t$ (blue) and $L=400$, and the analytical expression, Eq.~\eqref{backscatter2} (solid red line and dashed purple line) with $\alpha$ a fitting parameter. The dotted line shows perfect conductance $\bar{K}e^2/h$, Eq.~\eqref{fixedPoint}, with $\bar{K}\approx 0.739$ and $\bar{K}\approx 0.940$ respectively. The arrows at the $t_\ell$-axes indicate the points $u_\ell=u_w$.}
\label{Fig7}
\end{figure}

The half-filled case, Eq.~\eqref{XXZ} with $\mu_j\equiv 0$, is the
easiest to analyze for two reasons. First, the homogeneous spinless
fermion model is integrable for all interaction strengths $U_j\equiv
U$ and chemical potentials $\mu_j\equiv\mu$. However, only for $\mu=0$
(density $\langle n_j\rangle =1/2$) can the velocity of the elementary
excitations $u$ and the Luttinger liquid parameter $K$ be determined
in closed form
\begin{equation}
\label{vandK}
u= at\pi\frac{\sqrt{1-(U/2)^2}}{\arccos(U/2)},\quad K=\frac{\pi}{2[\pi-\arccos(U/2)]}\,.
\end{equation}
These results are valid in the critical regime $-2<U\leq 2$ where the
low-energy properties of the model are described by Luttinger liquid
theory. Second, we found in Ref.~\onlinecite{Sedlmayr2012a} that also
the criterion for perfect conductance at an abrupt junction is
particularly simple in this case. In general, each local perturbation
in the chain leads to an oscillating backscattering
$\sim\lambda\e^{-\im 2k_F x}\Psi_+^\dagger \Psi_-$ where $\Psi_\pm$
are the left- and right-moving fermion fields and $k_F$ is the Fermi
wavenumber with $k_F=\pi/2$ in the half-filled case. The scattering
amplitude $\lambda$ in the half-filled case takes the simple form
\begin{equation}
\label{ampl}
\lambda\propto \sum_j \e^{-\im2k_F j}(u_{j+1}-u_j)\,.
\end{equation}
While this amplitude averages to zero in the bulk of the lead and the
bulk of the wire, it is nonzero exactly at the boundary with 
\begin{equation}
\label{ampl2}
\lambda\propto u_\ell -u_w\,.
\end{equation}
For all interaction strengths in the critical regime we therefore
obtain a powerful and simple prediction for perfect conductance, i.e.,
conductance across a junction without any backscattering: {\it The
conductance is perfect if the velocity of excitations in the lead
$u_\ell$ exactly matches the velocity of excitations $u_w$ in the
interacting quantum wire.} If the condition
\eqref{ampl2} is fulfilled, then the conductance across a junction of a semi-infinite wire with LL parameter $K_\ell$ and a semi-infinite wire with LL 
parameter $K_w$ is given by\cite{Sedlmayr2012a,Sedlmayr2014a}
\begin{equation}
\label{fixedPoint}
G = \frac{e^2}{h}\bar{K}\quad\mbox{with}\quad \bar{K}=\frac{2K_\ell
K_w}{K_\ell + K_w}\,.
\end{equation}
For a noninteracting lead this reduces to $\bar{K}=2K_w/(1+K_w)$. In
Fig.~\ref{Fig7} we provide a numerical test of this
prediction comparing the conductance from QMC with the theoretically
predicted value for ideal conductance
\eqref{fixedPoint} if $u_\ell=u_w$.
Furthermore, we show that the conductance away from the fixed point is
well fitted by the second order perturbative result
\begin{equation}
\label{backscatter2}
G(T) = \frac{e^2}{h}\left[\bar{K}-\underbrace{\alpha(u_\ell-u_w)^2}_{\lambda^2}\left(\frac{T}{T_K}\right)^{2\bar{K}-2}\right]
\end{equation}
where the amplitude $\alpha$ is a 
free parameter. Note that for repulsive interactions, $K_w<1$,
backscattering is always relevant and increases in the limit $T\to
0$.  The conductance curve shown in Fig.~\ref{Fig7} is then
expected to become singular and will approach zero everywhere except at the conducting fixed
point.

\subsection{Away from half-filling}
Next, we want to study the conductance in the spinless fermion model
\eqref{XXZ} with a constant but nonzero chemical potential,
$\mu_j\equiv\mu\neq 0$. In this case, the condition $u_\ell=u_w$ for
perfect conductance across a junction no longer holds. Instead, we can
calculate the backscattering amplitude $\lambda$ only to lowest order
in the interaction and find\cite{Sedlmayr2014a}
\begin{eqnarray}
\label{backmu}
\lambda&\approx& \frac{a}{2\pi}\left(\frac{t_\ell}{\sin[ak_{F\ell}]}+\frac{U_\ell}{\pi}-\frac{t_w}{\sin[ak_{Fw}]}-\frac{U_w}{\pi}\right)\nonumber \\
&& -\frac{a\mu}{4\pi}\left(\cot[ak_{F\ell}] - \cot[ak_{Fw}]\right)\,.
\end{eqnarray}
Surprisingly, the scattering amplitude in lowest order is
real. Numerically, we have found that this seems to be the case even
for strong interactions. As a consequence, it should still always be
possible to find a conducting fixed point. Eq.~\eqref{backscatter2}
continues to describe the scaling of the conductance if the proper
backscattering amplitude is used. The LL parameter $K_w$ for the
interacting wire can no longer be written down in closed
form. However, it is possible to determine $K_w$ to high accuracy by
numerically solving integral equations obtained by Bethe ansatz.\cite{Essler2005} In a
previous paper, Ref.~\onlinecite{Sedlmayr2014a}, we have shown that
for every chemical potential $\mu$ it is possible to induce a sign
change in the Friedel oscillations near the junction by tuning the
parameters of the lead and wire. Since the Friedel oscillations are
linear in the backscattering amplitude $\lambda$ (see
Ref.~\onlinecite{Sedlmayr2014a} and Sec.~\ref{analyticSec}) this shows
that one can change the sign of $\lambda$ thus providing an indirect
proof for a conducting fixed point where $\lambda=0$.

\begin{figure}
\includegraphics[width=0.95\columnwidth]{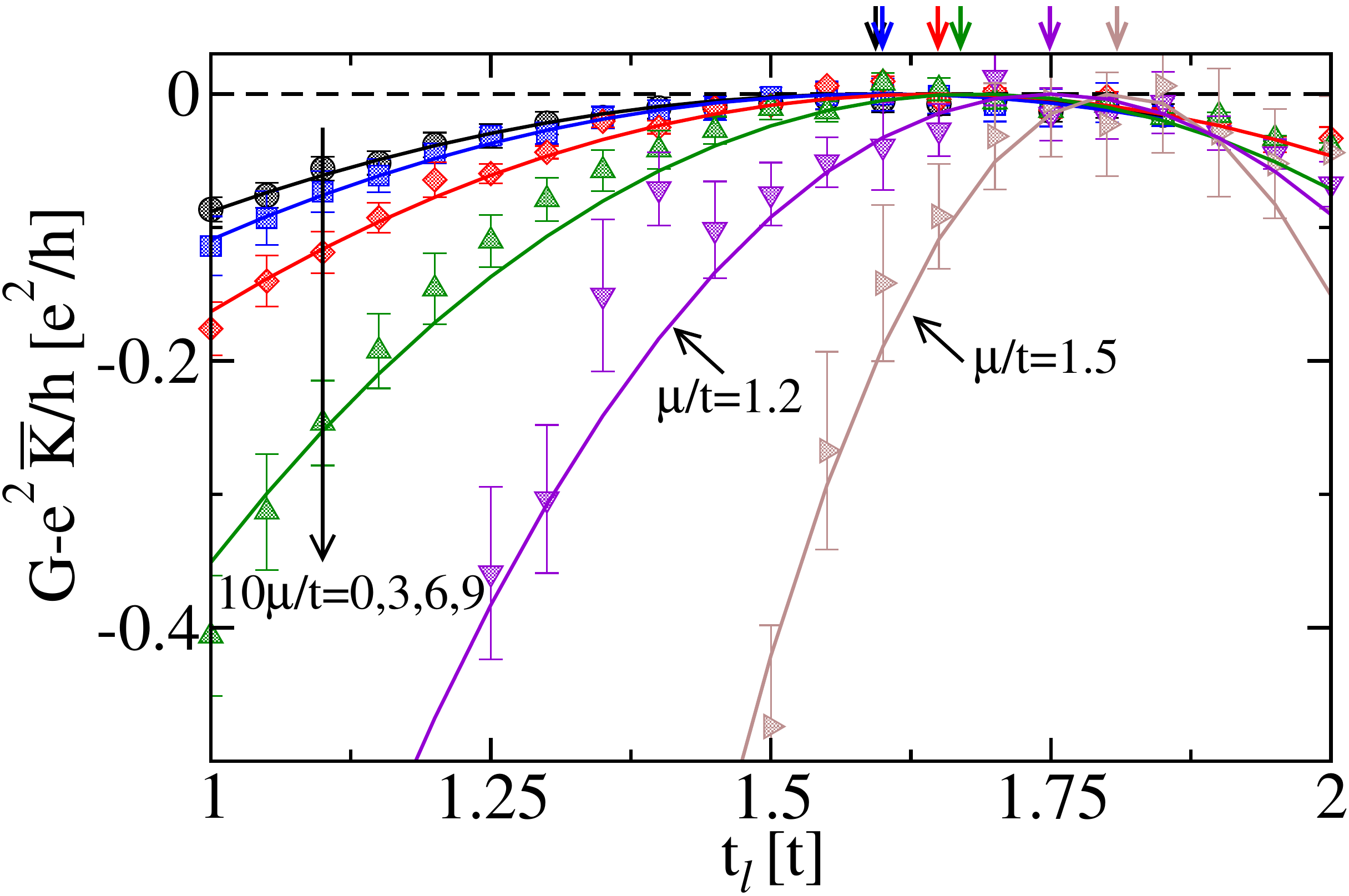} 
\caption{(Color online) Conductance $G-\bar{K}e^2/h$ for the inhomogeneous spinless chain \eqref{XXZ} with parameters $t_w=t$, $U_w=1.8t$, $U_\ell=0$, $\beta t =50$, and $L=400$, for different chemical potentials $\mu$. The density varies between $n\approx 0.51$ ($\mu/t=0.1$) and $n\approx 0.75$ ($\mu/t=1.5$). The dashed line indicates perfect conductance. The solid lines show fits to $-A(u_\ell-B)^2$ with $A$ and $B$ fitting parameters, see Eq.~\eqref{backmu}, the most general correction consistent with the model. The arrows indicate the points when $u_\ell=B$.}
\label{Fig8}
\end{figure}

Here we want to show the existence of conducting fixed points away
from half-filling directly. In Fig.~\ref{Fig8} we
present QMC data for the conductance across a junction of a lead and
an interacting quantum wire for various spatially constant chemical
potentials.
Note that we plot the measured conductance $G$ minus the ideal
conductance without backscattering given by $\bar{K}e^2/h$, i.e.~the
zero line in the plot indicates perfect conductance where
backscattering at the junction is absent. For all chemical potentials
shown, the curves indicate the existence of a perfectly conducting
fixed point. As expected based on the lowest order result of the
backscattering amplitude, Eq.~\eqref{backmu}, the position of the
fixed point shifts as a function of chemical potential.

\section{The inhomogeneous Hubbard model}\label{analyticSec}

While the spinless case is the easiest to analyze theoretically and
nicely demonstrates the existence of nontrivial perfectly conducting
fixed points for abrupt lead-wire junctions, its value as a realistic
model to describe experiments on quantum wires is limited. While one
could potentially spin polarize electrons in strong magnetic fields
making them effectively spinless, the typical experimental setup will
involve spinful electrons. As a next step, we therefore want to
generalize the investigation of perfectly conducting fixed point to
the Hubbard model \eqref{Hubbard}. We will concentrate here on the experimentally most
common case without magnetic fields $B_j\equiv 0$. The Hubbard model
then possesses a $SU(2)$-spin symmetry which fixes the spin Luttinger
liquid parameter to $K_s\equiv 1$.

In the following we will first present the low-energy effective theory
for an inhomogeneous Hubbard chain and then compare this theory with
QMC data for the conductances across a lead-wire junction.

\subsection{Luttinger liquid theory}
The homogeneous Hubbard model at low energies in the critical regime
where both spin- and charge excitations are gapless can be described
by an effective quadratic bosonic theory, the Luttinger liquid. In the
following we assume that we can generalize this effective theory
directly to the inhomogeneous case. Such an approach where only a
narrow band of states near the Fermi momenta are kept is certainly
justified if the hopping and interaction parameters as well as the
chemical potentials in lead and wire are close enough so that  
backscattering is weak and only states close to the Fermi momenta will
be mixed. In the following, we implicitly assume that we are in such a
limit. For large inhomogeneities at the junctions 
only numerical data can clarify if the Luttinger liquid
theory results still holds qualitatively.

The lead-wire junction at low energies is described by the effective
Hamiltonian $H=H_{\rm q}+H_{\rm bs}$ (see App.~\ref{app_bos} for details)
where
\begin{equation}\label{hq}
H_{\rm q}=\frac{1}{2}\int\ud x\sum_{\nu=c,s}\left(\frac{u^\nu_x}{K^\nu_x}(\partial_x\phi_i)^2+u^\nu_xK^\nu_x(\partial_x\tphi_i)^2\right)\,
\end{equation}
describes the bosonic modes which obey the commutation relations
$[\phi_\nu(x),\Pi_\eta(x')]=\im\delta_{\nu\eta}\delta(x-x')$ for $\nu,\eta=c,s$ with
$\Pi_\nu(x)=\partial_x\tilde{\phi}_\nu(x)$ a conjugate momentum. The spin
($\nu=s$) and charge ($\nu=c$) velocities and Luttinger parameters,
$u^\nu_x$ and $K^\nu_x$, respectively completely characterize the
systems low energy properties. We focus again on the case of a sharp
jump where we have two different regions with $u^\nu_{x< 0}=u^\nu_{\ell}$
and $u^\nu_{x>0}=u^\nu_{w}$. Provided the two boundaries of the wire are
far enough apart this is sufficient to characterize the required
properties of the system.

Additionally we have local backscattering terms at the junctions,
\begin{eqnarray}\label{ham_bs}
H_{\rm bs}&\approx&\lambda_R\cos[\sqrt{2\pi}\phi_c(0)]\cos[\sqrt{2\pi}\phi_s(0)]\\
&&+\lambda_I\sin[\sqrt{2\pi}\phi_c(0)]\cos[\sqrt{2\pi}\phi_s(0)]\,.\nonumber
\end{eqnarray}
Here $\lambda_R$ denotes the real part and $\lambda_I$ the imaginary part
of the scattering amplitude. Note that the
$\sin[\sqrt{2\pi}\phi_s(0)]$ term is forbidden by the $SU(2)$ symmetry
in the case without magnetic fields which is considered here. To lowest order
in the interaction one can calculate the backscattering coefficients
and we find
\begin{eqnarray}\label{back}
\lambda_R&=&\frac{1}{\pi}\left(\frac{v^1_\ell}{\sin^2[ak_{F\ell}]}-\frac{v^1_w}{\sin^2[ak_{Fw}]}-\frac{2aU_\ell}{\pi}+\frac{2aU_w}{\pi}\right.\nonumber\\
&&\quad+a\mu_\ell\cot[ak_{F\ell}]-a\mu_w\cot[ak_{Fw}]\bigg)\,\textrm{, and}\\\nonumber
\lambda_I&=&\frac{2aU_\ell}{\pi^2}\cot[ak_{F\ell}]-\frac{2aU_w}{\pi^2}\cot[ak_{Fw}]
\end{eqnarray}
with $v^1_{\ell,w}$ being renormalized Fermi velocities defined in
App.~\ref{app_bos}, and $k_{F\ell,w}$ being the Fermi momenta in the
lead and the wire, respectively. This result generalizes
Eq.~\eqref{backmu} to the spinful case. Importantly, the scattering
amplitude $\lambda$ is no longer real. This means that now, in
general, two separate conditions have to be fulfilled to make the
backscattering amplitude zero. Here we want to concentrate on a
non-interacting lead, $U_\ell=0$. In this case the imaginary part of
the backscattering amplitude is given by $\lambda_I\sim
U_w\cot[k_{Fw}a]$. In order for $\lambda_I$ to vanish either (i)
$U_w=0$ or (ii) $\cos[k_{Fw}a]=0$. The first case is not of interest
to us and leads for $\mu_\ell=\mu_w$ to the trivial fixed point of a
non-interacting homogeneous system. The second possibility implies
that the wire is half-filled, $k_{Fw}=\pi/2$ and $\mu_w=0$. Then
Eq.~\eqref{back} implies that one can find a point where $\lambda_R=0$
for any set of hopping and interaction wire parameters, $t_w$ and
$U_w$. This would make the backscattering in the half-filled
inhomogeneous Hubbard model analogous to the spinless case considered
before. However, even in the absence of backscattering at the
junction, the umklapp scattering term
\begin{equation}
\label{Umklapp}
H_U\sim U\int \ud x \cos(\sqrt{8\pi}\phi_c(x))
\end{equation}
is non-oscillating and relevant for repulsive interactions leading to
the charge gap \eqref{charge_gap} at half-filling. Therefore only the
spin sector can show ideal conductance at a non-trivial fixed point
for half-filling. Note that for attractive interactions the charge
sector remains gapless while a gap develops in the spin sector. In
this case, $K_w^c>1$ so that backscattering at the junction is always
irrelevant leading to perfect charge conductance. Away from
half-filling, on the other hand, Eq.~\eqref{back} suggests that
non-trivial conducting fixed points do not exist at all. However, it
is important to stress that this analysis is based on an expansion of
the scattering amplitude to lowest order in the Hubbard
interaction. Only numerical calculations can clarify if this result
also holds qualitatively for strong inhomogeneities.

The calculation of the conductance in a lead-wire-lead Hubbard system
for weak backscattering is a straightforward generalization of the
result in the spinless case, Eq.~\eqref{backscatter2}. 
In the model considered here the only backscattering
present is the $2k_F$ spin-conserving backscattering given by
Eq.~\eqref{ham_bs} in bosonized form. In general, also other sources
of backscattering---including processes which include a spin
flip---can be present at the boundary which could lead to different
backscattering amplitudes for charge and spin. The change of the
conductance as a function of temperature (energy scale) is determined
by the scaling dimension of the boundary operator. This scaling
dimension is found from the renormalization group (RG) equation
\begin{equation}
\frac{1}{\lambda}\frac{\ud \lambda}{\ud \ln T}=\bar K_c/2 + \bar K_s/2 - 1\,,
\end{equation}
where
\begin{equation}
\frac{1}{\bar{K}_\nu}=\frac{1}{2}\left[\frac{1}{K^\nu_{\ell}}+\frac{1}{K^\nu_{w}}\right]\,,
\end{equation}
and
\begin{equation}
\frac{1}{\bar{u}_\nu}=\frac{1}{2}\left[\frac{1}{u^\nu_\ell}+\frac{1}{u^\nu_w}\right]\,,
\end{equation}
with $\nu=c,s$. For a lead-wire system, i.e., a single junction between
a lead and a wire the ideal conductance in the absence of
backscattering now reads
\begin{equation}
g_0^{c,s}=\frac{2K_\ell^{c,s}K_w^{c,s}}{K_\ell^{c,s}+K_w^{c,s}}
\left\{\begin{array}{cc}
2\frac{e^2}{h}&\textrm{ for charge,}\\&\\
\frac{\mu_B^2}{2h}&\textrm{ for spin,}\end{array}\right.
\end{equation}
where again
$K^s_\ell=K^s_w\equiv 1$ for $SU(2)$ invariant models, although 
finite size and temperature can give significant logarithmic corrections.\cite{Soeffing2013}
Finally one finds for the differential conductance\cite{Furusaki1996}
\begin{equation}
g^{c,s}=g_0^{c,s}-\alpha_{c,s}\left(\frac{|\lambda|}{\bar
u_{c,s}}\right)^2\left(\frac{T}{T_K}\right)^{\bar K_{c}+\bar{K_s}-2}\,,
\end{equation}
where $T_K$ is the characteristic
temperature scale set by the backscattering strength and $\alpha_{c,s}$ are
constants. 

\subsection{Conductances from QMC}
As for the spinless case we will, in the following, use the SSE code
to calculate the conductances across an abrupt lead-wire junction
described by the inhomogeneous Hubbard model \eqref{Hubbard}. Based on
the analysis of the lowest order result for the backscattering
amplitude
\eqref{back} we might expect that the half-filled, particle-hole symmetric case is different
from any other generic filling. We will therefore discuss this case
separately.

\subsubsection{The half-filled case}
\begin{figure}[h!]
\includegraphics[width=0.99\columnwidth]{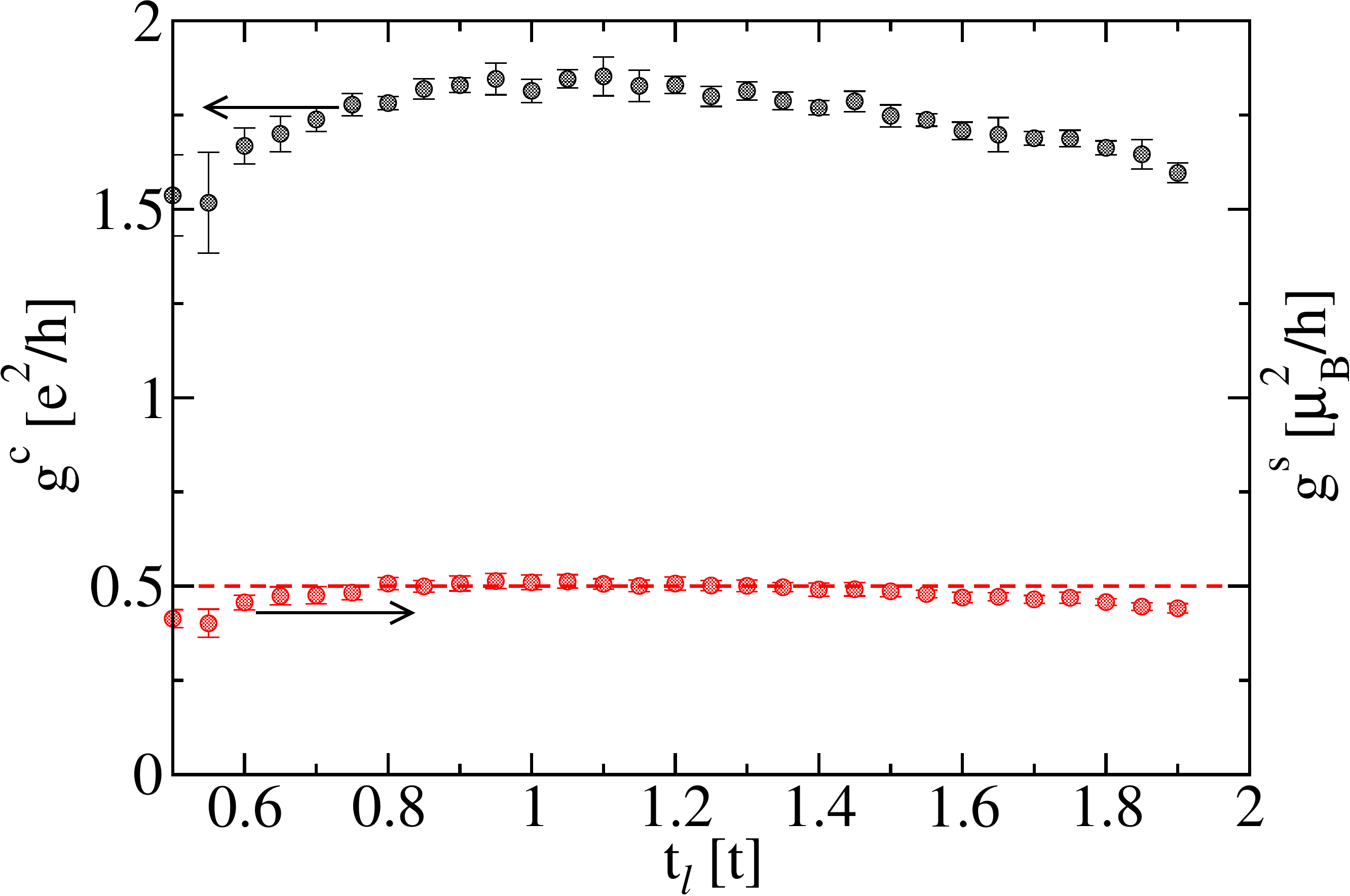} 
\caption{(Color online) Charge and spin conductance for the inhomogeneous half-filled Hubbard model for a chain with $L=150$, $t_w=t$, $\beta t=25$, and $U_w/t=1$. We find ideal spin conductance for $t_\ell\approx 1$.}
\label{Fig9}
\end{figure}

For half-filling the backscattering amplitude \eqref{back} to lowest
order in the Hubbard interaction is real. If this also holds for
stronger interactions then we might expect to be able to find a
non-trivial conducting fixed point for any set of wire parameters by
changing the hopping $t_\ell$ in the non-interacting lead. At this
fixed point we expect ideal spin conductance while the charge
conductance will become zero in the thermodynamic limit due to the
relevant umklapp scattering term in the bulk, Eq.~\eqref{Umklapp}. In
Fig.~\ref{Fig9} we exemplary show results for the case $U_w=1$.
%
% The spin conductance indeed reaches its ideal value $g^s=\mu_B^2/2h$ in
% a region around $t_\ell\approx 1$. The maximum is, however, quite
% broad so that it is not possible to determine the fixed point precisely. The charge conductance also shows a maximum in the same region. Similar to the homogeneous case shown in Fig.~\ref{Fig5} the conductance is nonzero only because the chain length in the numerical simulations is small compared to the length scale set by the exponentially small charge gap \eqref{charge_gap}. In the thermodynamic limit, the charge conductance will vanish for all hopping parameters $t_\ell$.
%
The spin conductance indeed reaches its ideal value $g^s=\mu_B^2/2h$ in
a region around $t_\ell\approx 1$. The maximum is, however, quite
broad so that it is not possible to determine the fixed point precisely. The charge conductance also shows a maximum in the same region. Similar to the homogeneous case shown in Fig.~\ref{Fig5} the conductance is nonzero only because the temperature
in the numerical simulations is large compared to the exponentially small
charge gap \eqref{charge_gap}. In the low-temperature limit, the charge conductance will vanish for all hopping parameters $t_\ell$.

\subsubsection{Away from half-filling}
\begin{figure}
\includegraphics[width=0.95\columnwidth]{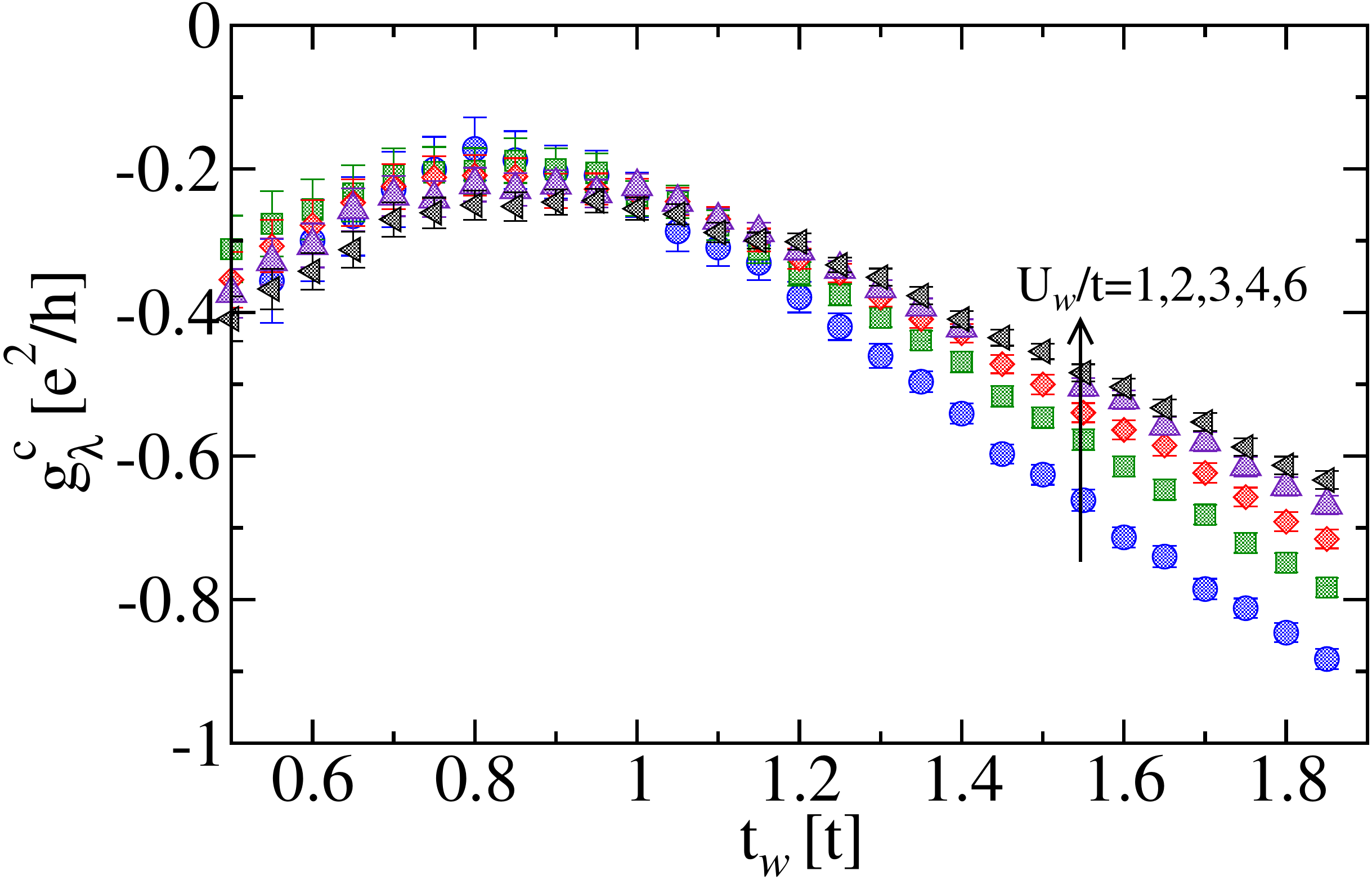}
\caption{(Color online) Relative differential charge conductance $g_\lambda^c=g^c-g_0^c$ for different values of $U_w$ as a function of $t_w$ in an inhomogeneous system of length $L=150$ with $t_\ell=t$, $\beta t=25$ and constant filling $n=1/4$ along the chain.}
\label{Fig10}
\end{figure}

\begin{figure}
\includegraphics[width=0.95\columnwidth]{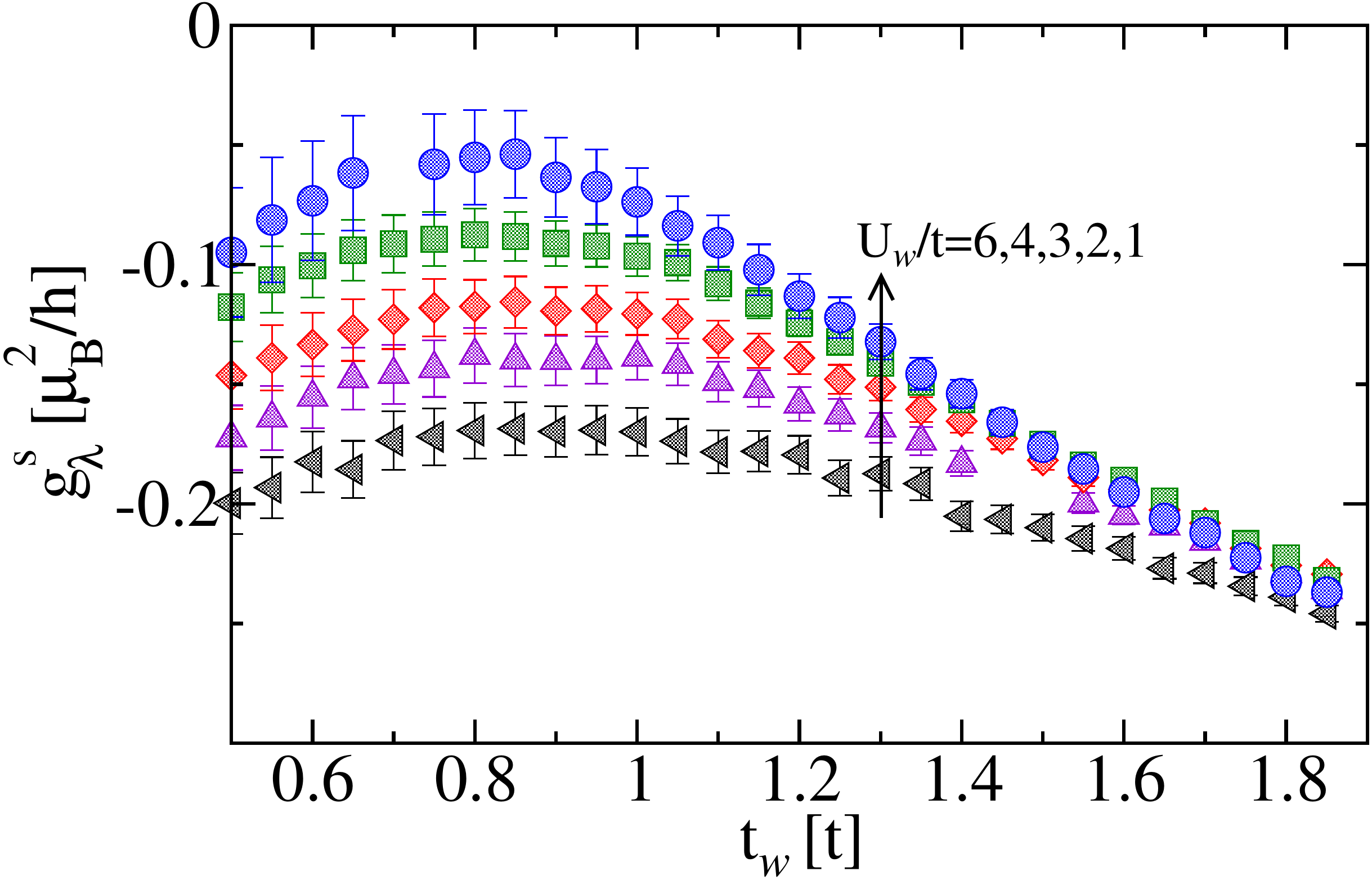} 
\caption{(Color online) Relative differential spin conductance $g_\lambda^s=g^s-g_0^s$ for different values of $U_w$ as a function of $t_w$ in an inhomogeneous system of length $L=150$ with $t_\ell=t$, $\beta t=25$ and constant filling $n=1/4$ along the chain.}
\label{Fig11}
\end{figure}

Away from half-filling the analysis of the lowest order result for the
scattering amplitude suggests that non-trivial conducting fixed points
do not exist.  Checking all possible parameter combinations in the
lead and in the wire numerically is not feasible, so this statement
cannot be explicitly shown.  However, it is possible to numerically
test several different cases by keeping the parameters in the
non-interacting lead fixed and vary both interaction and hopping
strength in the wire. Here the density across the junction is kept
constant at a generic value $n=1/4$ by choosing the chemical
potentials $\mu_\ell$ and $\mu_w$ accordingly.  In Figs.~\ref{Fig10}
and \ref{Fig11} we plot the relative conductances
$g_\lambda^{c,s}=g^{c,s}-g_0^{c,s}$ so that $g_\lambda^{c,s}=0$ would
correspond to a conducting fixed point. Note that we vary $t_w$ here
so that the Luttinger parameter $K_w^c$ and therefore $g_0^c$ is
different for each point shown in Figs.~\ref{Fig10} and \ref{Fig11}.
For both the spin and the charge conductance we see that the value for ideal conductance 
$g_\lambda^{c,s}=0$ is
never reached.   This is in contrast to the spinless case in Fig.~\ref{Fig8}, 
where for a given value 
on one side it was possible to achieve perfect conductance by just varying a single
parameter on the other side.
While this does not prove the conjecture---based on the
lowest order results for the scattering amplitude---that non-trivial
conducting fixed point do not exist in the spinful case away from
half-filling, it shows that the spinful case is different from the
spinless case.

\subsection{Friedel oscillations}
The inhomogeneity at a lead-wire junction leads to Friedel
oscillations in the local density which are proportional to the
backscattering amplitude $\lambda$.\cite{Sedlmayr2012a,Sedlmayr2014a,Rommer2000a}
 Calculating these oscillations for
small inhomogeneities by field theory and comparing the results with
QMC data is therefore an alternative way to study backscattering at
the junction. In Ref.~\onlinecite{Sedlmayr2014a} we have shown that
such an analysis can be used to find conducting fixed points in the
spinless case. In the following, we will generalize the field theory
for the Friedel oscillations to the spinful case and exemplary
compare the result with numerical data. 

The bosonized density operator for spinful fermions is given by
\begin{eqnarray}
\label{density_op}
n(x)&=&\nonumber n_{0}(x)-\frac{1}{\sqrt{\pi}}\partial_x\phi_c(x)\\&&\nonumber+\frac{\textrm{const.}}{\pi a}\sin[2k_{Fx}^*x+\sqrt{2\pi}\phi_c(x)]\cos[\sqrt{2\pi}\phi_s(x)]\,.
\end{eqnarray}
The oscillating contribution to the density is therefore obtained by
the following expectation value
\begin{equation}\label{rhoalt}
\rho_{\rm alt}(x)\propto\left\langle\sin[2k_{Fx}^*x+\sqrt{2\pi}\phi_c(x)]\cos[\sqrt{2\pi}\phi_s(x)]\right\rangle
\end{equation}
which has to be calculated with the full bosonized Hamiltonian
including the backscattering term. Here $k_{Fx}^*$ is the Fermi
momentum which can be found in the grand canonical setting from the
bulk density $\rho_x\equiv\langle n_{0}(x)\rangle=k_{Fx}^*/\pi$ which
is temperature dependent. In the following we will calculate the
Friedel oscillations
\eqref{rhoalt} to first order in the effective backscattering
coefficient $\lambda$. For this we require the following integral:
\begin{eqnarray}
\tau(x)&\equiv& 4\int_0^\beta\ud\tau\prod_{\nu=c,s}\left\langle
\e^{\im\sqrt{2\pi}\left[\phi_\nu(x,0)-\phi_\nu(0,\tau)\right]}\right\rangle\nonumber\\
&=&\int_0^\beta\ud\tau \prod_{\nu=c,s}\e^{\pi\left[G_\nu(x,0;\tau)-G_\nu(0,0;0)\right]}\,.
\end{eqnarray}
The Green's function $G_i(x,y,\tau)$ is defined in Appendix \ref{G_Appendix} and
can be obtained as a direct generalization of the spinless case, see
Ref.~\onlinecite{Sedlmayr2014a}. Note that in the spinful case the
integral consists of a product of a spin and a charge vertex operator
correlation function and can therefore no longer be evaluated
exactly. The integral can be cast into the following form
\begin{eqnarray}
\tau(x)&=&\prod_{\nu=c,s}\left(\frac{4\pi Ta}{u^\nu_{x}}\right)^{\frac{\bar{K}_\nu}{2}}\left(\frac{u^\nu_{x}}{2\pi aT} \sinh\left[\frac{2\pi Tx}{u^\nu_{x}}\right]\right)^{-\frac{K^\nu_{x}}{2}}\nonumber\\
&&\times\int_0^\pi\frac{\ud y}{\pi T}\prod_{\nu=c,s}\left[z_\nu+\sqrt{z_\nu^2-1}\cos[y]\right]^{-\frac{\bar{K}_\nu}{2}}\,,
\end{eqnarray}
with
\begin{equation}
z_\nu\equiv \coth\left[\frac{2\pi Tx}{u^\nu_x}\right]\,.
\end{equation}
The final result for the Friedel oscillations to first order in the
backscattering is
\begin{equation}\label{spin_dense}
\rho_{\rm alt}(x)\propto-\big[\lambda_R\sin[2k_{Fx}^*x]+\lambda_I\cos[2k_{Fx}^*x]\big]\tau(x)\,.
\end{equation}

\begin{figure}
\includegraphics[width=0.95\columnwidth]{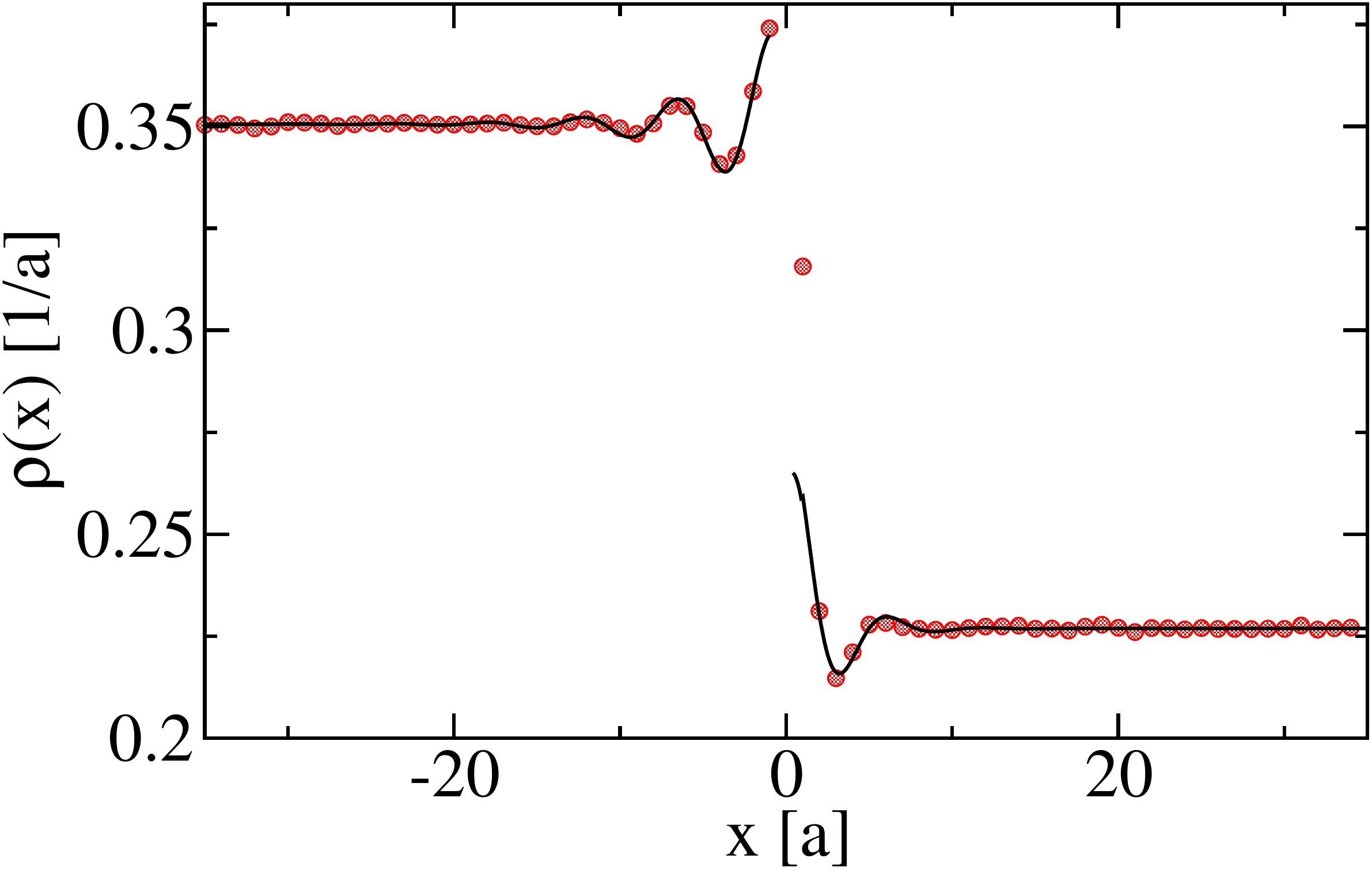}
\caption{(Color online) Local density at a lead-wire junction. Symbols denote the QMC results, the solid line denotes the Luttinger liquid result formula \eqref{spin_dense}. The numerical data are obtained for an inhomogeneous Hubbard model with length $L=150$, hoppings $t_\ell=t$, $t_w=0.8t$, wire interaction $U_w=2t$, and inverse temperature $\beta t_w=25$. The chemical potentials are $\mu_\ell=1.7t$, $\mu_w=1.7t$. Error bars for the QMC data are smaller than the symbols.}
\label{Fig12}
\end{figure}

In Fig.~\ref{Fig12} we compare the field theory formula
\eqref{spin_dense} with QMC results for the local density near the
boundaries of a lead-wire junction.
Sites with $x<0$ represent the non-interacting lead, sites with $x>0$
the interacting quantum wire. The bulk densities in the bulk of the
lead and the wire can be calculated by Bethe ansatz and are consistent
with the numerical data. To fit the alternating part of the local
density, both the position of the scattering center as well as the
amplitude of the oscillations are used as fitting parameters. The
obtained fit describes the data very well, showing that the field
theory description of the inhomogeneous system is working although the
inhomogeneity in the considered example is not small. A more detailed
study of the Friedel oscillations across the full parameter space of
the inhomogeneous Hubbard model \eqref{Hubbard} can, in principle, be
used to search for conducting fixed points. Similar to the
conductance, however, it is nearly impossible to show that non-trivial
conducting fixed points do not exist away from half-filling because of
the large parameter space which needs to be covered.

\section{Conclusions}\label{sec_conclusion}

Quantum wires---electrically conducting wires with diameters in
the nanometer range in which quantum effects strongly influence the
transport properties---offer insights into fundamental questions of
many-body physics as well as possible avenues to new electronic
devices. It is therefore important to develop numerical and analytical
tools to investigate the properties of such systems.

In this paper we have studied, in particular, the simplest quantum
wire device: an interacting quantum wire contacted by non-interacting
leads. Contrary to most previous studies, we model the lead-wire
junction microscopically and include electron scattering at the
junction. The latter is ignored in the most commonly used field
theoretical description of this setup where the junctions between
leads and quantum wire are assumed to be perfectly adiabatic. Our
microscopic approach starts from the opposite limit of a sharp
junction leading to models of inhomogeneous tight-binding chains where
parameters such as the hopping amplitude, the chemical potential, and
the screened Coulomb interactions abruptly change on the scale of the
lattice spacing.

To numerically investigate lead-wire junctions we have generalized a
quantum Monte Carlo algorithm based on the stochastic series expansion
technique which has been used previously for homogeneous
systems.\cite{Louis2003} This method allows us to calculate response
functions in imaginary time. We calculate the linear response to an infinitesimal drop
in electric or magnetic field. After a Fourier
transformation to discrete Matsubara frequencies we have shown that at
sufficiently low temperatures a reliable extrapolation to zero
frequency is possible, giving access to the charge and spin conductance
near zero temperature. To test the validity and accuracy of this
approach we have studied different homogeneous and inhomogeneous
setups where the conductances are known exactly. In all those test
cases we have found very good agreement of the numerical data with the
exact results, establishing this method as a reliable tool to study
quantum wire devices.

As a first application, we have studied the conductance across a
lead-wire junction in a spinless fermion system. In two previous
publications,\cite{Sedlmayr2012a,Sedlmayr2014a} we have predicted by
field theoretical means that non-trivial perfectly conducting fixed
points exist despite the inhomogeneity of the system on the scale of
the lattice spacing. At these fixed points the amplitude of the
relevant backscattering process exactly vanishes. For the half-filled
spinless fermion system we have predicted this to happen when the
velocities of the excitations in lead and wire exactly
match. Previously, we have only been able to provide indirect
numerical evidence for this fixed point by studying Friedel
oscillations and autocorrelations near the junction. Here we have
directly calculated the conductance and shown that the result near the
fixed point can be well fitted by the field theory formula requiring
only a single fitting parameter. Next, we have also studied the
conductance in inhomogeneous spinless fermion wires away from
half-filling. In this case, field theory predicts that conducting
fixed points still exist, however, the condition for perfect
conductance is no longer a simple velocity matching. We have verified
this prediction here numerically as well; values close to perfect
conductance are obtained for all fillings investigated.

While spinless fermions are easiest to study by field theory, the
spinful case is the experimentally more relevant one. To study whether
or not non-trivial conducting fixed points still exist once the spin
degree of freedom is included we have analyzed the inhomogeneous
Hubbard chain without magnetic field using bosonization. This analysis
provided evidence for a fundamental difference to the spinless case:
while the amplitude of the relevant backscattering process is always
real for spinless fermions it is complex, in general, for the spinful
case. For the $SU(2)$ symmetric inhomogeneous Hubbard chain, in
particular, we find to lowest order in the Hubbard interaction that
the imaginary part of the backscattering amplitude only vanishes at
half-filling (particle-hole symmetric case). If we conjecture that
this holds to all orders in the interaction, then non-trivial
conducting fixed points only exist for the half-filled
system. Numerically, we have been able to show the existence of a
conducting fixed point at half-filling for the inhomogeneous Hubbard
model where the spin conductance takes it ideal value $\mu_B^2/2h$
while the charge conductance will vanish in the thermodynamic limit
due to the charge gap induced for repulsive interactions by a relevant
bulk umklapp scattering term. On the other hand, a non-trivial fixed
point was not found for several lead-wire setups away from
half-filling.

There seem to be therefore two main setups in which these conducting
fixed points---described by a rather unusual boundary conformal field
theory\cite{Sedlmayr2014a}---can possibly be investigated
experimentally. On the one hand, one might consider a quantum wire of
spin polarized electrons which is effectively described by a spinless
fermion model. On the other hand, it might be possible to use a
spinful quantum wire with a low-energy band structure which can be
tuned to a particle-hole symmetric filling by a gate electrode. In
both cases the field theory predicts that for a sufficiently sharp
junction a non-trivial conducting fixed point should be accessible by
tuning the effective bandwidths and chemical potentials of the
leads. For the half-filled spinful model, in particular, a fixed point
with perfect spin conductance can be found for repulsive interactions
while perfect charge conductance is expected for attractive
interactions with backscattering at the junction being always
irrelevant in the latter case.

Finally, we note that the experiment described in Ref.~\onlinecite{KamataKumada}
has recently been analyzed using the bosonic model \eqref{hq} but
without the local backscattering term
\eqref{ham_bs}.\cite{Perfetto2014,Calzona2015a} In these studies the
authors find backscattering of a wavepacket injected into the lead at
a lead-wire junction. We want to stress that this result is not in
contradiction to the results presented here. While a wavepacket is
indeed scattered at the junction in an inhomogeneous Luttinger model
\eqref{hq} even without a single electron backscattering term
\eqref{ham_bs} being present, the conductance will be ideal in this case as has already been
stressed in Ref.~\onlinecite{Safi1995}.

\acknowledgments
J.S. acknowledges support by the Natural Sciences and Engineering
Research Council (NSERC, Canada) and by the Deutsche
Forschungsgemeinschaft (DFG) via Research Unit FOR 2316.
This research was supported by the DFG via Transregio 49,  Transregio 173, and Transregio 185 (S.E.~and D.M.).
Support for this research at Michigan State
University (N.S.) was provided by the Institute for Mathematical and
Theoretical Physics with funding from the office of the Vice President
for Research and Graduate Studies. We are grateful for computation
time at AHRP.

\appendix

\section{Luttinger liquid theory}\label{app_bos}

The low energy behavior of the Hamiltonian, Eq.~\eqref{Hubbard}, can
be described as a Luttinger liquid.\cite{Giamarchi2004} Here we extend
our analysis to a broader class of interactions that also includes the
nearest neighbor interactions $V_{j,j+1}$. We will set $\hbar=1$ everywhere in this appendix. The interacting
Hamiltonian now reads
\begin{eqnarray}\label{fullhamiltonian}
H_I&=&\sum_{j,\sigma\sigma'}\bigg[U_j:\dpsi_{\sigma j}\psi_{\sigma j}::\dpsi_{\sigma' j}\psi_{\sigma' j}:\\\nonumber&&
\qquad+V_{j,j+1}:\dpsi_{\sigma j}\psi_{\sigma j}::\dpsi_{\sigma' j+1}\psi_{\sigma' j+1}:\bigg]\,.
\end{eqnarray} 
Normal ordered operators are given by $:\dpsi_j\psi_j:=\dpsi_j\psi_j-\langle0|\dpsi_j\psi_j|0\rangle$, with $|0\rangle$ the ground state. To simplify matters we consider a spin-independent, $SU(2)$ symmetric interaction. The low-energy theory does however remain valid for a spin-dependent interaction provided the interaction is spin conserving. The derivation of the spatially inhomogeneous Luttinger liquid theory follows closely the standard homogeneous derivation\cite{Safi1995,Maslov1995} but special care must be taken to include the local scattering at the boundary.\cite{Sedlmayr2012a,Sedlmayr2014a}

We assume that we can linearize the dispersion near the Fermi momenta $k_{Fx}$ into left and right moving particles via the ansatz 
\begin{equation}\label{lin_ansatz}
 \frac{\psi_{\sigma j}}{\sqrt{a}}=\psi_\sigma(x)=\sum_{\alpha=\pm}\e^{\im\alpha k_{Fx}x}\psi_{\sigma\alpha}(x)
\end{equation}
where the Fermi momenta are given by
\begin{equation}\label{mu}
\mu_{x}=-2t_x\cos\left[ak_{Fx}\right]\,,
\end{equation}
with $a$ the lattice spacing and $x=aj$.
The $\alpha=+$ and $\alpha=-$ indices denote the right and left moving electrons respectively.
After linearization we have for the noninteracting part of Eq.~\eqref{Hubbard}, taking the continuum limit,
\begin{eqnarray}\label{linHam}
H_{0}&=&-\int\ud x\sum_{\alpha=\pm,\sigma}t_x\left[e^{\im ak_{Fx}}\dpsi_{\sigma\alpha}(x)\partial_{x}\psi_{\sigma\alpha}(x)+\textrm{H.c.}\right]\nonumber\\
&&-\int\ud x\sum_{\alpha=\pm,\sigma}\left[2t_xe^{-i\alpha k_{Fx}(2x+a)}+\mu_{x} e^{-2\im\alpha k_{Fx}x}\right]\nonumber\\
&&\qquad\times\dpsi_{\sigma\alpha}(x)\psi_{\sigma\bar{\alpha}}(x)\,.
\end{eqnarray}
We have defined $\bar{\alpha}=-\alpha$.
The contribution of the final two lines can be neglected in a homogeneous system, but will here still contribute near the boundary where $t_x$ and $\mu_x$ can sharply vary.
Using Eq.~\ref{lin_ansatz} the interaction $H_I$ can be decomposed into parallel and perpendicular spin components, which in the usual nomenclature\cite{Giamarchi2004} are written as
\begin{eqnarray}
\label{g-terms}
H_2&=&\sum_{\sigma,\alpha}\int\ud x\left[
\frac{g^{2\perp}_x}{2}\rho_{\sigma \alpha}\rho_{\bar{\sigma}\bar{\alpha}}+
\frac{g^{2\parallel}_x}{2}\rho_{\sigma \alpha}\rho_{\sigma\bar{\alpha}}\right],\nonumber\\
H_4&=&\sum_{\sigma,\alpha}\int\ud x\frac{g^{4}_x}{2}\rho_{\sigma \alpha}\rho_{\sigma \alpha}\,\textrm{, and}\\
H_1&=&\sum_{\sigma, \alpha}\int\ud x\left[
\frac{g^{1\perp}_x}{2}\psi^\dagger_{\sigma \alpha}\psi^\dagger_{\bar{\sigma}\bar{\alpha}}\psi_{\bar{\sigma}\alpha}\psi_{\sigma \bar{\alpha}}
-\frac{g^{1\parallel}_x}{2}\rho_{\sigma \alpha}\rho_{\sigma \bar{\alpha}}\right]\,.\nonumber
\end{eqnarray}
Here we have suppressed the spatial indices and defined the local right and left mover density $\rho_{\sigma\pm}=\psi^\dagger_{\sigma\pm}\psi_{\sigma\pm}$.
Note that the $g^{1\parallel\sigma}_x$ process has been rewritten from its natural form to resemble a density-density interaction, however the final $g^{1\perp}$ process can not be formulated as a density-density interaction. It corresponds to a  two particle backward scattering process. This is at best marginal, and will be neglected here. Umklapp scattering processes, when they are important, lead to a charge gap, these are discussed in the main text. In addition there are scattering terms in $H_I$ which originate with the inhomogeneity of the interaction which renormalize the backscattering in Eq.~\eqref{linHam}.\cite{Sedlmayr2012a,Sedlmayr2014a}

We introduce bosonic fields related to the particle density,
\begin{equation}
\rho_{\sigma \alpha}(x)=-\frac{1}{\sqrt{2\pi}}\partial_x\phi_{\sigma \alpha}\,,
\end{equation}
which obey the commutation relations
\begin{eqnarray}
\left[\phi_{\sigma\alpha}(x),\phi_{\sigma'\alpha}(x')\right]&=&-\frac{\im}{2}(\alpha\delta_{\sigma\sigma'}\sgn(x-x')
+\im\sigma^y_{\sigma\sigma'})\,\textrm{, and}\nonumber\\
\left[\phi_{\sigma\alpha}(x),\phi_{\sigma'\bar{\alpha}}(x')\right]
&=&-\frac{\im\alpha}{2}(\delta_{\sigma\sigma'}+\im\sigma^y_{\sigma\sigma'})\,.
\end{eqnarray}
The vertex operator is
\begin{equation}
\psi_{\sigma\alpha}(x)=\frac{1}{\sqrt{2\pi a}}e^{\im\alpha\sqrt{2\pi}[\phi_{\sigma\alpha}(x)]}\,.
\end{equation}
The Hamiltonian can be reformulated as a quadratic Hamiltonian in these bosonic fields, $H_{\rm q}$, and local scattering terms $H_{\rm bs}$.

Firstly the quadratic part of the bosonic Hamiltonian is
\begin{widetext}
\begin{equation}
H_{\rm q}=\frac{1}{8\pi}\int\ud x
\begin{pmatrix}
\partial_x\phi_{\uparrow+}\\\partial_x\phi_{\uparrow-}\\\partial_x\phi_{\downarrow+}\\\partial_x\phi_{\downarrow-}
\end{pmatrix}^T
\begin{pmatrix}
4\pi v_{Fx}+2g^4_x & g^{2\parallel}_x-g^{1\parallel}_x & g^4_x & g^{2\perp}_x \\
g^{2\parallel}_x-g^{1\parallel}_x & 4\pi v_{Fx}+2g^4_x & g^{2\perp}_x & g^4_x \\
g^4_x & g^{2\perp}_x & 4\pi v_{Fx}+2g^4_x & g^{2\parallel}_x-g^{1\parallel}_x \\
g^{2\perp}_x & g^4_x & g^{2\parallel}_x-g^{1\parallel}_x & 4\pi v_{Fx}+2g^4_x
\end{pmatrix}
\begin{pmatrix}
\partial_x\phi_{\uparrow+}\\\partial_x\phi_{\uparrow-}\\\partial_x\phi_{\downarrow+}\\\partial_x\phi_{\downarrow-}
\end{pmatrix}\,.
\end{equation}
\end{widetext}
The unrenormalized velocity is $v_{Fx}=2at_{x}\sin[ak_{Fx}]$.
We make two unitary transformations which suffice to diagonalize $H_{\rm q}$.
The first is
$\phi_{\sigma\pm}=[\phi_{\sigma}\mp \tphi_{\sigma}]/\sqrt{2}$. The new fields obey $[\phi_\sigma(x),\Pi_{\sigma'}(x')]=\im\delta_{\sigma\sigma'}\delta(x-x')$, with the conjugate momentum $\Pi_\sigma(x)=\partial_x\tphi_\sigma(x)$.
The second transformation is to rotate to the spin-charge representation: $\phi_{c/s}=[\phi_{\uparrow}\pm\phi_{\downarrow}]/\sqrt{2}$ (and similar for the $\tphi_\sigma(x)$ fields). These obey similar commutation relations $[\phi_\nu(x),\Pi_{\mu}(x')]=\im\delta_{\nu\mu}\delta(x-x')$, with the conjugate momentum $\Pi_\nu(x)=\partial_x\tphi_\nu(x)$ and $\nu=c,s$.

We now have the diagonal representation
\begin{equation}
H_{\rm q}=\int\ud x[\partial_x\Phi(x)]^T\mathbf{M}\partial_x\Phi(x)\,.
\end{equation}
where $\mathbf{M}(x)$ is the diagonal matrix
\begin{equation}
\label{M_SpinCharge}
\mathbf{M}(x)=\frac{1}{2}
\diag\left(\frac{u^c_x}{K^c_x},\frac{u^s_x}{K^s_x},u^c_x K^c_x,u^s_x K^s_x\right)\,,
\end{equation}
and $[\Phi(x)]^T=(\phi_{c},\phi_{s},\tphi_{c},\tphi_{s})$.
Here $K^s_x$ and $K^c_x$ are the spin and charge Luttinger
parameters, and $u^s_x$ and $u^c_x$ are the renormalized spin and
charge velocities. These parameters are functions of the interaction
strengths and Fermi velocities, and to lowest order can be calculated
directly:
\begin{eqnarray}
  u^c_x&\approx&v_{Fx}+\frac{g^{4}_x}{\pi}\,,\nonumber\\
  u^s_x&\approx&v_{Fx}\,,\\
  K^c_x&\approx&1-\frac{1}{2\pi}\frac{g^{2\parallel}_x-g^{1\parallel}_x+g^{2\perp}_x}{v_{Fx}}\,,\textrm{ and}\nonumber\\\nonumber
  K^s_x&=&1\,.\label{ks}
\end{eqnarray}
At the non interacting $SU(2)$ symmetric point the Luttinger
parameters are given by $K^s_x=K^c_x=1$.

Collecting terms from both Eqs.~\eqref{linHam} and \eqref{fullhamiltonian}, and using
\begin{equation}
\dpsi_{\sigma\alpha}\psi_{\sigma\bar{\alpha}}(x)=\frac{\im\alpha}{2\pi a}e^{-\im\alpha\sqrt{4\pi}\phi_\sigma(x)}\,,
\end{equation}
the local scattering at the boundary is
\begin{eqnarray}
H_{\rm bs}&=&\frac{1}{2\pi\im a}\sum_{\substack{x=ja\\j\in\mathbb{Z}}}\sum_\sigma\e^{-2\im k_{Fx}x-\im\sqrt{4\pi}\phi_\sigma(x)}\\&&\nonumber
\times\left[\frac{e^{-\im ak_{Fx}}u^1_{x}}{\sin[ak_{Fx}]}-\frac{2a\im U_x}{\pi}+a\mu_{x}\right]+\textrm{H.c.}\,,
\end{eqnarray}
which has been written again as a sum. We have used the renormalized velocity,
\begin{equation}
u^1_{x}=v_{Fx}+\frac{4aV^1_x}{\pi}\sin^2[ak_{Fx}]\,,
\end{equation}
calculated to lowest order. We have also defined $U_x=U_j$ and $V^1_x=V_{j,j+1}$ in the continuum limit with $x=aj$.

On performing the sum only local contributions from the discontinuities at the boundary survive. In the case of a single junction as in Table \ref{table_parameters} the necessary sum can be approximately evaluated as
\begin{equation}
\sum_{\substack{x=ja\\j\in\mathbb{Z}}}\e^{-2\im k_{Fx}x}F_x\approx\frac{\im F_\ell\e^{\im ak_{F\ell}}}{\sin[ak_{F\ell}]}-\frac{\im F_w\e^{\im ak_{Fw}}}{\sin[ak_{Fw}]}
\end{equation}
with $F_x$ varying as the parameters in Table \ref{table_parameters}.
The backscattering Hamiltonian then becomes
\begin{eqnarray}
H_{\rm bs}&\approx&\Re\lambda\cos[\sqrt{2\pi}\phi_c(0)]\cos[\sqrt{2\pi}\phi_s(0)]\\\nonumber
&&+\Im\lambda\sin[\sqrt{2\pi}\phi_c(0)]\cos[\sqrt{2\pi}\phi_s(0)]\,,
\end{eqnarray}
with
\begin{eqnarray}
\lambda&=&\frac{u^1_{\ell}}{\pi\sin^2[ak_{F\ell}]}-\frac{u^1_{w}}{\pi\sin^2[ak_{Fw}]}
\\\nonumber&&
+\frac{1}{\pi^2}\left(a\pi\mu_{\ell}-2\im aU_\ell\right)\left(\cot[ak_{F\ell}]+\im\right)\\\nonumber&&
-\frac{1}{\pi^2}\left(a\pi\mu_{w}-2\im aU_w\right)\left(\cot[ak_{Fw}]+\im\right)\,.
\end{eqnarray}
The full Luttinger liquid description of the model is given by the Hamiltonian $H=H_{\rm q}+H_{\rm bs}$.

\section{The Green's function}\label{G_Appendix}

For the spinful Hamiltonian $H_{\rm q}$, Eq.~\eqref{hq}, we can calculate the charge and spin Green's functions:
\begin{equation}
G_\nu(x,y;\tau)=\langle\phi_\nu(x,0)\phi_\nu(y,\tau)\rangle\,.
\end{equation}
These satisfy the differential equation
\begin{equation}\label{gmat}
\bigg[\frac{\omega_m^2}{K^\nu_xu^\nu_x}-\frac{\partial}{\partial x}\bigg(\frac{u^\nu_x}{K^\nu_x}\frac{\partial}{\partial x}\bigg)\bigg]G_{\nu,m}(x,x')=\delta(x-x')\,,
\end{equation}
where
\begin{eqnarray}
G_\nu(\tau,x,x')=T\sum_me^{i\omega_m\tau}G_{\nu,m}(x,x')\,,
\end{eqnarray}
for $\omega_m=2\pi m T$ with $m\in\mathbb{Z}$.
We have set the lattice spacing here to $a=1$.  We introduce here the function $\mathcal{L}_{x,y}$ which is equal to $1$ when $x$ and $y$ are in the same `region', and $0$ when they are not.
Eq.~\eqref{gmat} can be solved giving\cite{Maslov1995,Sedlmayr2014a}
\begin{eqnarray}\label{tg}
G_{\nu,m}(x,y)
&=& \frac{\bar{K}_\nu}{2|\omega_m|}e^{-\frac{|\omega_n||x|}{u^\nu_x}-\frac{|\omega_n||y|}{u^\nu_y}}
\\ \nonumber&&+\frac{\mathcal{L}_{x,y}K^\nu_x}{2|\omega_m|}\left[e^{-\frac{|\omega_n||x-y|}{u^\nu_x}}-e^{-\frac{|\omega_n|(|x|+|y|)}{u^\nu_x}}\right],\nonumber
\end{eqnarray}
and therefore
\begin{eqnarray}
G_\nu(x,y;\tau)&=&
-\frac{\bar{K}_\nu}{2\pi}\ln\left|\sinh\left[\pi T\left(\frac{|x|}{u^\nu_x}+\frac{|y|}{u^\nu_y}-i\tau\right)\right]\right|
\\ \nonumber&&+\frac{\mathcal{L}_{x,y}K^\nu_x}{2\pi}\ln\left|\frac{\sinh\left[\pi T\left(\frac{|x|}{u^\nu_x}+\frac{|y|}{u^\nu_y}-i\tau\right)\right]}{
\sinh\left[\pi T\left(\frac{|x-y|}{u^\nu_x}-i\tau\right)\right]}\right|\,,\\ \nonumber&&
\end{eqnarray}
for the required Green's function.

% \bibliography{../../../../Papers/references}


\end{document}